\documentclass{article}

\usepackage{arXiv}

\usepackage[utf8]{inputenc} 
\usepackage[T1]{fontenc}    
\usepackage{hyperref}       
\usepackage{url}            
\usepackage{booktabs}       
\usepackage{amsfonts}       
\usepackage{nicefrac}       
\usepackage{microtype}      
\usepackage{lipsum}		
\usepackage{graphicx}
\usepackage{doi}

\usepackage{natbib}
\setcitestyle{authoryear,round} 
\hypersetup{
    colorlinks=true,
    linkcolor=blue,      
    citecolor=blue,      
    urlcolor=blue        
}

\title{Modelling hydroelastic flexure of arbitrarily shaped ice shelves forced by long ocean waves}

\author{
    T.K.~Papathanasiou \\
    National Technical University of Athens, Greece \\
    \texttt{t\_papathanasiou@mail.ntua.gr} 
    \And
    L.G.~Bennetts \\
	The University of Melbourne, Australia \\
	\texttt{luke.bennetts@unimelb.edu.au} 
	\And
	M.H~Meylan \\ 
	University of Newcastle, Australia \\
	\texttt{mike.meylan@newcastle.edu.au}
}

\date{}

\renewcommand{\shorttitle}{Flexure of arbitrarily shaped ice shelves}

\hypersetup{
pdftitle={Flexure of arbitrarily shaped ice shelves},
pdfsubject={q-bio.NC, q-bio.QM},
pdfauthor={Papathanasiou~et~al.},
pdfkeywords={First keyword, Second keyword, More},
}

\usepackage{amsmath}

\usepackage{enumerate}

\usepackage{overpic,rotating}

\newcommand{\beq}{\begin{equation}}
\newcommand{\eeq}{\end{equation}}
\newcommand{\bea}{\begin{eqnarray}}
\newcommand{\eea}{\end{eqnarray}}
\newcommand{\bean}{\begin{eqnarray*}}
\newcommand{\eean}{\end{eqnarray*}}

\newlength{\mylinelength}
\newlength{\mydashlength}
\newlength{\mydashspace}
\newlength{\mychainlengthA}
\newlength{\mychainlengthB}
\newlength{\mychainspace}
\newlength{\mylinethickness}
\usepackage{pifont}
\ifdefined\showmylabels
 \usepackage[inline]{showlabels}
 
\fi

\usepackage[normalem]{ulem}
\usepackage{soul}
\usepackage{tcolorbox}
\ifdefined\shownewold
   \newcommand{\del}[1]{{\color{red}\sout{#1}}}
\else
   \newcommand{\del}[1]{\ignorespaces}
\fi
\ifdefined\shownewold
   \newcommand{\deleqn}[1]{\\\del{\parbox{\textwidth}{#1}}}
\else
   \newcommand{\deleqn}[1]{\ignorespaces}
\fi
\ifdefined\shownewold
   \newcommand{\new}[1]{{\color[rgb]{0,0,1}#1}}
\else
   \newcommand{\new}[1]{#1}
\fi
\ifdefined\shownewold
   \newcommand{\old}[1]{{\color[rgb]{0.5,0.5,0.5}#1}}
\else
   \newcommand{\old}[1]{\ignorespaces}
\fi
\ifdefined\shownewold
   \newcommand{\delref}[1]{\del{\parbox{\figwidth\columnwidth}{#1}}}
\else
   \newcommand{\delref}[1]{\vspace{-12pt}}
\fi
\ifdefined\shownewold
   \newcommand{\lu}[1]{{\color[rgb]{1,0,1} #1}}
   \newcommand{\luchange}[2]{\del{#1}\new{#2}}
   \newcommand{\luleft}[1]{{\color[rgb]{1,0,1}$\longleftarrow$#1}}
   \newcommand{\luright}[1]{{\color[rgb]{1,0,1}#1$\longrightarrow$}}
\else
   \newcommand{\lu}[1]{\ignorespaces}
   \newcommand{\luchange}[1]{\ignorespaces}
   \newcommand{\luleft}[1]{\ignorespaces}
   \newcommand{\luright}[1]{\ignorespaces}
\fi
\usepackage{tikz,pgfplots}
\usetikzlibrary{calc,patterns,decorations.pathmorphing,decorations.markings,decorations.pathmorphing,angles,quotes,arrows,arrows.meta}
\tikzset{snake it/.style={decorate, decoration=snake}}

\newlength{\figurewidth}
\newlength{\figwidth}
\newlength{\figureheight}

\usepackage{xcolor}
\definecolor{darkgreen}{rgb}{0,0.55,0}
\definecolor{midgreen}{rgb}{0,0.8,0.2}
\definecolor{magenta}{rgb}{1,0,1}
\definecolor{purple}{rgb}{0.5,0,0.5}
\definecolor{darkorange}{rgb}{1,0.55,0}
\definecolor{maroon}{rgb}{0.5,0,0}
\definecolor{olive}{rgb}{0.5,0.5,0}
\definecolor{midgrey}{rgb}{0.5,0.5,0.5}
\definecolor{lightgrey}{rgb}{0.75,0.75,0.75}
\definecolor{matlabblue}{rgb}{0,0.447,0.741}
\definecolor{matlabred}{rgb}{0.85,0.325,0.098}
\definecolor{lightblue}{rgb}{0,0.5,1}
\definecolor{darkgrey}{rgb}{0.25,0.25,0.25}
\definecolor{teal}{rgb}{0,0.5,0.5}
\definecolor{navy}{rgb}{0,0,0.5}
\definecolor{goldenrod}{rgb}{0.85,0.6,0.1}

\usepackage{amsmath}
\usepackage{amssymb}
\usepackage{bm}
\usepackage{siunitx}

\newcommand{\mathsfbi}[1]{\bm{\mathsf{#1}}}

\begin{document}
\maketitle

\begin{abstract}
	Flexure of Antarctic ice shelves under excitation from long ocean waves induces mechanical ice shelf stresses that amplify fractures and, hence, contribute to calving events. 
Here, a solution method is developed for a hydroelastic mathematical model of wave-induced ice shelf flexure, based on the conventional  theory of a Kirchoff-Love plate floating on shallow water under linearised conditions, but allowing wave forcing of ice shelves with variations in both horizontal dimensions, and where the ice shelves are of arbitrary shape, including non-uniform thickness. 
The method uses finite elements specifically designed for the high-order hydroelastic system, and a Dirichlet-to-Neumann map to bound the computational domain in the open ocean.
Following verification, the method is used to conduct novel studies on how the ice-shelf deflection is affected by the ice shelf shape, the incident wave direction and the proportion of the shelf that is grounded. 
The efficiency of the method allows the studies to be conducted over a broad frequency range, such that resonant responses are identified. 
\end{abstract}


%
\section{Introduction}
\label{sec:intro}

\begin{figure}
  \centerline{
  \begin{tabular}{c @{\hspace{20pt}} c}
       \includegraphics[height=6cm]{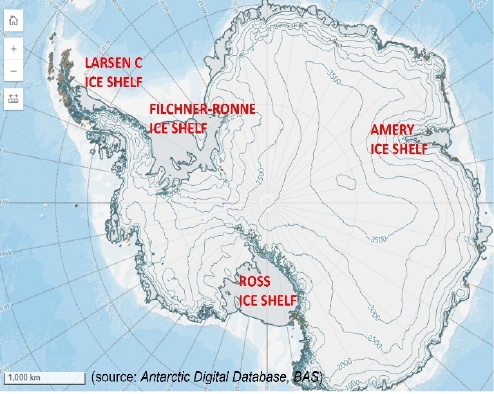}
       &  
       \includegraphics[height=6cm]{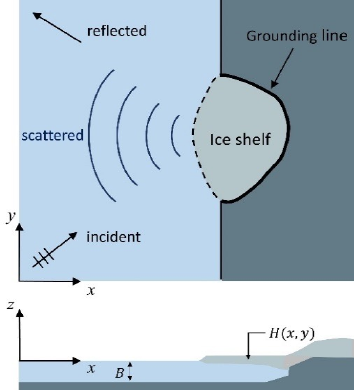}
  \end{tabular} 
  }
  \centerline{
  \includegraphics[width=0.99\linewidth]{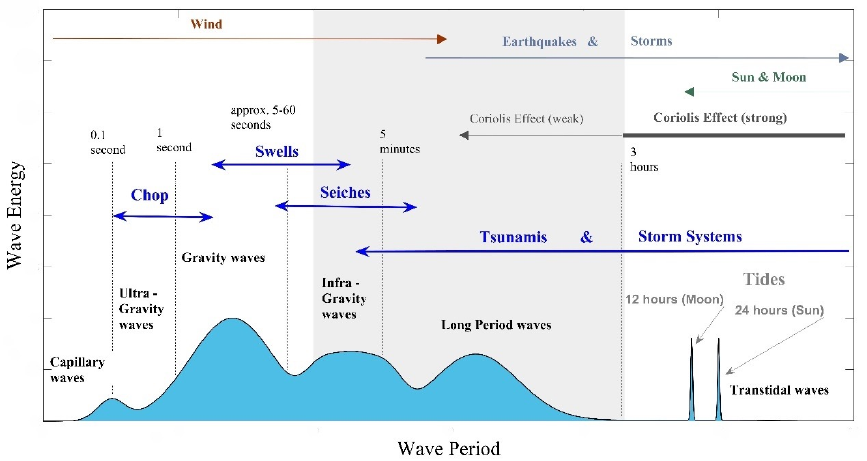}
  }
  \caption{Major Antarctic ice shelves (top-left). Geometry of the ice shelf scattering problem under long wave excitation. Ice shelf of variable thickness $H(x,y)$ over variable bathymetry $B(x,y)$ and clamped at the grounding line (top-right).
  Ocean wave energy spectrum and wave period bands. 
  The shaded area indicates the intended validity range for the model used in this study (bottom).} 
\label{fig:fig1}
\end{figure}

Glaciers that flow down to a coastline and onto the ocean surface form floating platforms known as ice shelves (or ice tongues) that enclose cavities of ocean water. Ice shelves are found in several polar regions, but they are largest and most prominent around Antarctica, where individual ice shelves can be the size of countries, and collectively they occupy over half the coastline (Figure~\ref{fig:fig1}, top-left). 
Antarctic ice shelves are fundamental components of the Southern Ocean and Antarctica \citep{bennetts2024closing}, and they play a major role in mitigating sea level rise by buttressing outflow of the Antarctic Ice Sheet \citep{noble2020sensitivity}. 
During the past decades, many Antarctic ice shelves have been losing mass to thinning and calving, which is impairing their buttressing ability \citep{depoorter2013calving,greene2022calving}. 
Certain ice shelves have suffered large and unprecedented calving events \citep{brunt2011antarctic,scambos2013climate,massom2018disintegration,arthur2021triggers,gomezfell2022parker,banwell2017calving,ochwat2023triggers,zhao2024longterm}. 

Ocean waves over a wide range of wave periods ($\approx 15$\,s--4\,h; Figure~\ref{fig:fig1}, bottom) have been implicated in calving events, including long swell \citep{banwell2017calving,massom2018disintegration,ochwat2023triggers,teder2025large}, infragravity waves \citep{bromirski2010transoceanic} and tsunami waves \citep{brunt2011antarctic,zhao2024longterm}.
The mechanism is the flexure of the ice shelves forced by ocean waves \citep{thiel1960gravimetric,williams1981flexural,robinson1992travelling,squire1994observations,macayeal2006transoceanic,cathles2009seismic,bromirski2010transoceanic,bromirski2015ross,bromirski2017tsunami,chen2018ocean,chen2019ross}, which induces stresses and strains that amplify fractures and promote calving \citep{holdsworth1978iceberg,bassis2012upper,lipovsky2018ice,bassis2024stability}.
The flexure is believed to  reach damaging levels only for ice shelves that have thinned and become heavily damaged, and, in the case of long swell, have lost their surrounding sea ice barriers \citep{massom2018disintegration,teder2022sea}.

Mathematical models have been developed to predict wave-induced ice shelf flexure \citep{bennetts_waves_2025}, dating back to the seminal work by \citet{holdsworth1978iceberg,holdsworth1981mechanism}.
The models typically use Kirchoff-Love elastic plate theory for the ice shelf, coupled with shallow-water theory for the sub-shelf water cavity and surrounding ocean, on the basis that wavelengths are much greater than the ice thickness and water depth. They are a class of ``hydroelastic'' models.
They are also typically linearized on the basis of small amplitude motions. 
Moreover, the models assume the timescales of the wave--ice shelf interactions are much faster than those of the glacial ice flow, such that the ice flow can be neglected. 
The models can be viewed as extended versions of the classic fluid mechanics problem of harbour oscillations excited by long ocean waves \citep[][ Chap.~5]{mei2005theory}, where a plate covers the harbour and there is no breakwater to constrain the width of the connection to the open ocean.
In analogy to the harbour problem, the ice shelf response spectrum is characterised by resonances, which have been hypothesised to cause the greatest damage to ice shelves \citep[][and others since]{holdsworth1978iceberg}.

The ice shelf problem is made challenging by the coupled water--ice interactions (that create a sixth-order hydroelastic partial differential equation system), the irregular bathymetry/ice shelf topography (that create varying coefficients), the large spatial span of ice shelves and underlying water cavities, and the coupling to the surrounding ocean (that create large computational domains; Figure~\ref{fig:fig1}, top-right).
Hence,
most models have been limited to only one horizontal dimension and use uniform ice thickness and water depth \citep{vinogradov1985oscillation,sergienko2010elastic}. 
Moreover, to calculate the resonant frequencies, artificial zero pressure or zero flux conditions have been applied along the water column below the shelf front, which reduces the calculations to finding eigenvalues and eigenvectors \citep{sergienko2013normal,meylan2017normal}. 

A connection to the open ocean creates a scattering problem, in which the ice shelf response to ocean wave forcing is calculated over a continuous frequency spectrum. 
In contrast to the eigenproblem, the scattering problem allows the degree of ice shelf flexure in response to a prescribed ocean wave forcing to be quantified.
Resonances appear as peaks in a response function 
\citep{papathanasiou2018resonances,ilyas2018timedomain,meylan2021swell,kalyanaraman2019shallow}
and are connected with the eigenproblem through the theory of complex resonances \citep{kalyanaraman2020coupled,bennetts2021complex,kalyanaraman2021icefem}. 
The models contain extensions to varying seabed profiles \citep{papathanasiou2018resonances,ilyas2018timedomain} and ice thickness profiles \citep{kalyanaraman2020coupled,meylan2021swell,bennetts2021complex,kalyanaraman2021icefem}, along with finite water depths 
\citep{ilyas2018timedomain,kalyanaraman2020coupled,bennetts2021complex} and 2D linear elasticity for the ice shelf \citep{kalyanaraman2020coupled,kalyanaraman2021icefem}.

Models with a second horizontal dimension allow for a class of resonant modes that are not uniform in the spanwise direction \citep[the $y$-direction in Figure~\ref{fig:fig1}, top-right;][]{papathanasiou2019hydroelastic}, along with other potentially important features, such as lateral ice shelf boundaries. 
However, they demand numerical solution methods, and, hence, have only been used in a couple of studies 
\citep{sergienko2017behavior,tazhimbetov2023simulation}. 
A ‘brute force’ commercial software has been used but requires millions of degrees of freedom and extensive computational resources, such that only a handful of simulations have been analysed and with only an approximate connection to the open ocean \citep{sergienko2017behavior}.
Therefore, tailor-made, high-fidelity numerical codes are needed to study ice shelf responses to ocean waves over broad wave frequency ranges and ice shelf configurations. 
A finite difference scheme has recently been proposed \citep{tazhimbetov2023simulation}, which reduces the degrees of freedom, but requires a mapping of the complex spatial domains to simpler ones to apply the finite difference stencils and incorporate boundary conditions. It also requires corrections at the boundary corners. 
The mathematically similar problem of sea ice cover over a harbour has also been considered \citep{li2021interaction}, using a boundary integral equation solution method that leverages on the standard assumptions for sea ice based on its thinness (centimetres to metres only) that the floating plate has zero draught and is of uniform thickness. 

In this paper, we approach the ice shelf problem using a nonconforming hydroelastic finite element triangle that captures 2D horizontal effects for ice shelves of general shape and avoids the need for domain mapping. 
So far, it has only been applied to the eigenproblem in which the sub-shelf water cavity is cut off from the open ocean by no pressure or no flux boundary conditions \citep{papathanasiou2019hydroelastic}.
We make the important next step in model development, by connecting the ice shelf/cavity region with the open ocean and solving the scattering problem in which the ice shelf responds to an incident wave. 
We employ the Dirichlet-to-Neumann boundary condition that allows the open water region to be constrained to a relatively small size, similar to several studies on Helmholtz scattering problems \citep{melenk2010convergence,mitsoudis2012helmholtz}.
We validate the method by comparing it to a simpler but more expensive alternative in which the Sommerfeld radiation condition is applied on the semi-circular boundary at a large distance from the ice shelf. 
We use this method to study how the response of the ice shelf is affected by its shape, its protrusion (whether it is more like an ice shelf or ice tongue) and the angle of incidence, focusing on the resonant responses. 

\section{Formulation of the scattering problem}

\begin{figure}
    \centerline{
    \includegraphics[width=0.75\linewidth]{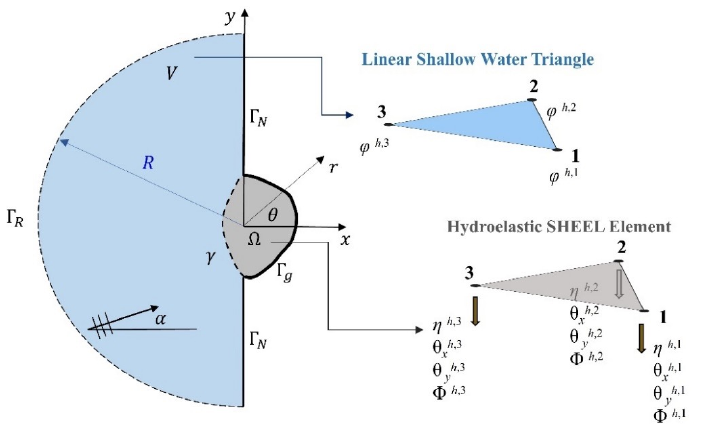}
    }
    \caption{Computational domain for the scattering problem. The ice-shelf/ice-shelf cavity domain $\Omega$ is discretised using the hydroelastic SHEEL element \citep{papathanasiou2019hydroelastic}, while the open ocean domain, $V$, is discretised with a Linear Shallow Water Triangle}
    \label{fig:comp_domain}
\end{figure}

Let locations in the horizontal plane be denoted by the Cartesian coordinate $(x,y)$.
The ice-shelf/sub-ice-shelf cavity domain is $(x,y)\in \Omega$, where $\Omega$ is bounded and open 
(Figure~\ref{fig:comp_domain}, left). 
The boundary of $\Omega$ includes the grounding line, $\Gamma_{\text{g}}$, and the interface with the surrounding ocean, $\gamma$. 
The ocean domain, $(x,y)\in V$, is a subset of the half-plane $x<0$, which is bounded by the coastline, $\Gamma_{\text{N}}$ (a subset of the $y$-axis). 
The coastline is assumed to be a straight, impermeable and, hence, perfectly reflecting wall for simplicity.
For computational purposes, $V$ is also bounded by an artificial interface, $\Gamma_R$, to the far-field,
$\sqrt{ x^{2} + y^{2}} \to \infty$.
In the scattering problem, the ice-shelf/sub-ice-shelf cavity system is forced by plane ocean waves travelling at a prescribed angle, $\alpha$, with respect to the horizontal $x$-axis (Figure~\ref{fig:comp_domain}, left).

The modelling assumptions compatible with the wave-band of interest (approximately 15\,s to 4\,h, covering the cases of long swell, infragravity waves and tsunamis; shaded area in Figure~\ref{fig:fig1}, bottom) are as follows.
\smallskip
\begin{itemize}
    \item The water motion is irrotational, governed by linear shallow water theory, such that the fluid velocity, $\bm{u}$, is 
    \[
    \bm{u} 
    = \nabla \Phi,
    \]
    where $\Phi(x,y,t)$ is the velocity potential and $t$ denotes time.
    \smallskip
    \item Coriolis effects are neglected, based on previous findings that they only have an appreciable influence for wave periods greater than 3--4\,h \citep[][; see also Appendix~\ref{append:coriolis}]{papathanasiou2019hydroelastic}.
    \smallskip
    \item The ice shelf is a slender, elastic solid, subject to Kirchhoff–Love assumptions of thin plate bending with small deflection, such that
    \[
    \frac{H}{\sqrt{A}} \ll 1,
    \]
    where $H$ is the ice shelf thickness and $A$ is the ice shelf horizontal area.
    \smallskip
    \item The plate material is isotropic, satisfying Hooke’s law. 
    Its elastic modulus is $E > 0$ and its Poisson’s ratio is $\nu $. The values $\nu = 0.3$ and $E = 11$ GPa are chosen as standard, although the Young’s modulus value is debated \citep{liang2024pan}.
    \smallskip
    \item The water density, $\rho_w = 1027~\mathrm{kg~m^{-3}}$, ice shelf density, $\rho_i = 917~\mathrm{kg~m^{-3}}$, and gravitational acceleration, $g = 9.81~\mathrm{m~s^{-2}}$, are constants. 
\end{itemize}

In the open ocean domain, $V$, continuity and momentum conservation for the linearized shallow water model are combined to form a D'Alembert-type equation for the velocity potential, $\varphi(x,y,t)$, such that
\begin{equation}\label{eq:shallow_water}
\frac{1}{g} \frac{\partial^2 \varphi}{\partial t^2} - \nabla \cdot (B \nabla \varphi) = 0,
\end{equation}
where $B$ is the location of the bathymetry relative to the equilibrium free surface, which is assumed constant, so that $\nabla \cdot (B \nabla \varphi)=B\nabla^{2}\varphi$.
In the ice-shelf/cavity domain, $\Omega$, the system of governing equations for the ice shelf flexural displacement, $\eta(x,y,t)$, and the velocity potential, $\Phi(x,y,t)$, in the water cavity are 
\begin{equation}\label{eq:plate1}
\rho_i H \frac{\partial^2 \eta}{\partial t^2}
- \frac{\partial^2 M_{xx}}{\partial x^2}
- 2 \frac{\partial^2 M_{xy}}{\partial x \partial y}
- \frac{\partial^2 M_{yy}}{\partial y^2}
+ \rho_w g \eta
+ \rho_w \frac{\partial \Phi}{\partial t}
= 0
\end{equation}
\begin{equation}\label{eq:plate2}
\text{and}\quad
\frac{\partial \eta}{\partial t}
+ \nabla \cdot \left( [B - d] \nabla \Phi \right)
= 0,
\end{equation}
where $H(x,y)$ is the ice shelf thickness and $d(x,y) = {\rho_i H(x,y)}/{\rho_w}$ is the ice-shelf draft. 
For a thin Kirchhoff–Love plate, the bending moments and curvatures are related through Hooke’s law as \citep{timoshenko1959theory}
\begin{equation}\label{eq:hooke}
\mathsfbi{M}(\eta) =
\begin{bmatrix}
M_{xx} & M_{xy} \\
M_{yx} & M_{yy}
\end{bmatrix}
= -D(x,y)
\begin{bmatrix}
\displaystyle \frac{\partial^2 \eta}{\partial x^2} + \nu \frac{\partial^2 \eta}{\partial y^2} & (1 - \nu) \frac{\partial^2 \eta}{\partial x \partial y} \\
(1 - \nu) \frac{\partial^2 \eta}{\partial x \partial y} & \displaystyle \frac{\partial^2 \eta}{\partial y^2} + \nu \frac{\partial^2 \eta}{\partial x^2}
\end{bmatrix},
\end{equation}
where
\[
D(x,y) = \frac{E H^3(x,y)}{12(1 - \nu^2)}
\]
is the ice shelf flexural rigidity.

Time-harmonic solutions at a prescribed angular frequency, $\omega$, are assumed, such that (acknowledging the abuse of notation) 
\[
\eta(x,y,t) = \eta(x,y)e^{-i\omega t}, \quad \Phi(x,y,t) = \Phi(x,y)e^{-i\omega t}
\quad \text{and} \quad \varphi(x,y,t) = \varphi(x,y)e^{-i\omega t}.
\]
Using \eqref{eq:hooke}, the governing equations \eqref{eq:shallow_water},
\eqref{eq:plate1} and \eqref{eq:plate2}  become
\begin{align}
\label{eq:plate3_freq}
g\nabla \cdot (B \nabla \varphi) + {\omega^2} \varphi &= 0, && \text{in } V
\\[4pt] \label{eq:plate1_freq}
\mathcal{L}\eta + g(1 - \omega^2 \mu)\eta - i\omega\Phi &= 0, && \text{in } \Omega  
\\[4pt] \label{eq:plate2_freq}
\text{and}\quad
\nabla \cdot \left( [B - d]\nabla \Phi \right) - i\omega\eta &= 0, && \text{in } \Omega. 
\end{align}
In Eq.~\eqref{eq:plate1_freq}, $\mu(x,y) = d(x,y)/g$ and the
operator  for a Kirchhoff–Love plate, $\mathcal{L}$, is defined as
\begin{equation}
\mathcal{L} \eta = \nabla^2 (K \nabla^2 \eta) 
- (1 - \nu)\left( \partial_{xx} K \partial_{yy} \eta 
- 2 \partial_{xy} K \partial_{xy} \eta 
+ \partial_{yy} K \partial_{xx} \eta \right), 
\end{equation}
with $K(x,y) = D(x,y)/\rho_w$.

The velocity potential in the open ocean domain, $\varphi(x,y)$, can be decomposed into components due to the incident wave, $\varphi_{\text{inc}}(x,y)$, the reflected wave, $\varphi_{\text{ref}}(x,y)$ (due to the straight coastline), and the scattered wave field, $\varphi_{\text{sca}}(x,y)$ (the perturbation to the reflected field due to the ice shelf plus radiation due to its motion), such that
\begin{equation}\label{eq:varphi_decomp}
\varphi = \varphi_{\text{inc}} + \varphi_{\text{ref}} + \varphi_{\text{sca}} = e^{ik(x \cos \alpha + y \sin \alpha)} + e^{-ik(x \cos \alpha - y \sin \alpha)} + \varphi_{\text{sca}}, 
\end{equation}
where $k = \omega / \sqrt{gB}$ is the wave number. 
The reflected potential is given explicitly in the right-hand side of \eqref{eq:varphi_decomp}, leaving the scattered potential, $\varphi_{\text{sca}}$, as the principal unknown.

For the long-period wave regime, it is appropriate to set the normal bending moment and active shear force to be zero at the lateral boundary of the ice shelf exposed to the ocean \citep{bennetts2024thin}. 
These boundary conditions have the form \citep{papathanasiou2019hydroelastic}
\begin{equation}\label{eq:bend_shear}
M_{n}(\eta) = \boldsymbol{n}^{T} \mathsfbi{M}(\eta) \boldsymbol{n} = 0 \quad \text{and} \quad Q(\eta) + \nabla T(\eta) \cdot \boldsymbol{t} = 0 \quad \text{on } \gamma, 
\end{equation}
where $T(\eta) = \boldsymbol{n}^T \mathsfbi{M}(\eta) \boldsymbol{t}$ is the twisting moment and 
$Q(\eta) = \nabla \cdot \mathsfbi{M}(\eta) \boldsymbol{n}$ is the shear force, with $\boldsymbol{t}$ being the unit tangent vector on $\gamma$.
At the grounding line, clamped boundary conditions are applied to the ice shelf \citep{holdsworth1981mechanism}, such that
\begin{equation}\label{eq:clamp}
\eta = 0, \quad \theta_n = 0 \quad \text{and} \quad \theta_{t} = 0 \quad \text{on } \Gamma_{\text{g}}, 
\end{equation}
where $\theta_n$, $\theta_t$ are the normal and tangential rotations, respectively. Impermeability of the ice-shelf cavity wall at the grounding line implies
\begin{equation}\label{eq:cavity_wall}
\nabla \Phi \cdot \boldsymbol{n} = 0 \quad \text{on } \Gamma_{\text{g}}. 
\end{equation}
Impermeability of the continental boundaries leads to the boundary condition
\begin{equation}\label{eq:continent}
\nabla \varphi \cdot \boldsymbol{n} = 0 \quad \text{on } \Gamma_{\text{N}}. 
\end{equation}
The interface conditions at the ocean-exposed front of the ice shelf, expressing conservation of momentum and mass flow, are
\begin{equation}\label{eq:interface}
\Phi = \varphi \quad \text{and} \quad (B - d)\nabla \Phi \cdot \boldsymbol{n} = B \nabla \varphi \cdot \boldsymbol{n} \quad \text{on } \gamma. 
\end{equation}

On the computational domain boundary, $\Gamma_{R}$ (Figure~\ref{fig:comp_domain}, left), a Dirichlet-to-Neumann (DtN) radiation condition \citep{oberai1998implementation} is applied, 
such that
\begin{equation}\label{eq:DtN}
\left. \frac{\partial \phi}{\partial r} \right|_{r = R} = f(R, \theta) + \int_{\pi/2}^{3\pi/2} G(R, \theta, \vartheta) \phi(R, \vartheta) \, d\vartheta. 
\end{equation}
 Appendix~\ref{append:DtN} provides details of the derivation of the DtN condition, along with expressions for the kernel function, $G(R, \theta, \vartheta)$, and the forcing term, $f(R, \theta)$. 
As an alternative to \eqref{eq:DtN}, the Sommerfeld radiation condition,
\begin{equation}\label{eq:sommerfeld}
\nabla \varphi_{\text{sca}} \cdot \boldsymbol{n} - ik \varphi_{\text{sca}} = 0, 
\end{equation}
can be applied, but requires the radius, $R$, to be large compared to the ice-shelf dimensions and the wavelengths involved, so that more finite elements must be used, with increased computational cost. In contrast, the DtN condition \eqref{eq:DtN} is valid for any radius $R$ \citep{melenk2010convergence} (if it covers the vicinity of the ice shelf). 
However, truncation of the series representing the kernel function, $G(R, \theta, \vartheta)$, is required, while the nested computations involved will increase computational time. 
In the following, \eqref{eq:DtN} will be primarily used and \eqref{eq:sommerfeld} will be employed for verification purposes. 

\section{Monolithic nonconforming finite element method}

\subsection{Variational form}

Testing \eqref{eq:plate1_freq} with function $\chi$, employing the Green–Gauss theorem twice and using the homogeneous conditions \eqref{eq:bend_shear} and \eqref{eq:clamp}, gives
\begin{equation}\label{eq:3.1}
a(\chi,\eta) + \int_\Omega g(1 - \omega^2 \mu) \bar{\chi} \eta \, d\Omega - i\omega \int_\Omega \bar{\chi} \Phi \, d\Omega = 0, 
\end{equation}
where
\begin{align}
a(\chi,\eta) &= \int_\Omega \nu K \Delta \bar{\chi} \, \Delta \eta \, dxdy \notag \\
&\quad + \int_\Omega K(1 - \nu)\left( \partial_{xx} \bar{\chi} \, \partial_{xx} \eta + \partial_{yy} \bar{\chi} \, \partial_{yy} \eta + 2\partial_{xy} \bar{\chi} \, \partial_{xy} \eta \right) dxdy. \notag 
\end{align}
Similarly, testing \eqref{eq:plate2_freq} with function $-W$, using the Green–Gauss theorem and the zero-flux condition at the ice-shelf cavity wall \eqref{eq:cavity_wall}, gives
\begin{equation}\label{eq:3.2}
\int_\Omega (B - d) \nabla \bar{W} \cdot \nabla \Phi \, d\Omega + i\omega \int_\Omega \bar{W} \eta \, d\Omega - \int_\gamma \bar{W} (B - d) \nabla \Phi \cdot \boldsymbol{n} \, d\Gamma = 0. 
\end{equation}
Testing \eqref{eq:plate3_freq} with function $-w$, using the Green–Gauss theorem and the no-flux condition at the impermeable boundary $\Gamma_{\text{N}}$ \eqref{eq:continent}, gives
\begin{equation}\label{eq:3.3}
\int_V B \nabla \bar{w} \cdot \nabla \varphi \, dV - \frac{\omega^2}{g} \int_V \bar{w} \varphi \, dV + \int_\gamma B \bar{w} \nabla \varphi \cdot \boldsymbol{n} \, d\Gamma - \int_{\Gamma_R} B \bar{w} \nabla \varphi \cdot \boldsymbol{n} \, d\Gamma = 0.
\end{equation}

Summing \eqref{eq:3.1}--\eqref{eq:3.3} and using the interface conditions \eqref{eq:interface} to cancel the integrals on $\gamma$, the variational form is
\begin{equation}
S([\chi, W, w], [\eta, \Phi, \phi]) + i\omega \mathcal{A}([\chi, W], [\eta, \Phi]) = F(w), 
\end{equation}
with the sesquilinear functional
\begin{align}
S([\chi, W, w], [\eta, \Phi, \phi]) &= 
a(\chi, \eta) + \int_\Omega g(1 - \omega^2 \mu) \bar{\chi} \eta \, d\Omega + \int_\Omega (B - d) \nabla \bar{W} \cdot \nabla \Phi \, d\Omega \notag 
\\ \nonumber
&\quad 
+ \int_V \left(B \nabla \bar{w} \cdot \nabla \phi - \frac{\omega^2}{g} \bar{w} \phi\right) dV 
\\&\quad
- \int_{\Gamma_R} B \bar{w} \int_{\pi/2}^{3\pi/2} G(R,\theta,\vartheta) \phi(R,\vartheta) d\vartheta \, d\Gamma, 
\end{align}
the skew-symmetric form
\begin{equation}
\mathcal{A}([\chi, W], [\eta, \Phi]) = \int_\Omega (\bar{W} \eta - \bar{\chi} \Phi) \, d\Omega, 
\end{equation}
and the linear forcing functional
\begin{equation}
F(w) = \int_{\Gamma_R} B \bar{w} f(R, \theta) \, d\Gamma. 
\end{equation}
Note that $i\mathcal{A}([\eta, \Phi], [\chi, W]) = -i\mathcal{A}([\chi, W], [\eta, \Phi])$, and, therefore, $i\mathcal{A}([\chi, W], [\eta, \Phi])$ is a Hermitian form 
\citep{papathanasiou2019hydroelastic,karperaki2019optimized}. 
This indicates the associated terms correspond to hydroelastic coupling and not to energy dissipation mechanisms.

For an open and bounded set $\Omega \subset \mathbb{R}^2$, we use the standard notation $H^k(\Omega; \mathbb{C})$ to denote the Sobolev (Hilbert) space $W^{k,2}(\Omega; \mathbb{C})$ over the complex numbers. 
It is $L^2(\Omega; \mathbb{C}) \equiv H^0(\Omega; \mathbb{C})$ and $(\bar{u}, v)_\Omega$ denotes the standard $L^2$-inner product for $u, v \in L^2(\Omega; \mathbb{C})$ with induced norm $\|u\|_{0,\Omega} = \sqrt{(\bar{u}, u)_\Omega}$. 
The inner-product induced norm for $H^k(\Omega; \mathbb{C})$ is denoted as $\|u\|_{k,\Omega}$ and the corresponding seminorm $|u|_{k,\Omega}$, where for the multi-index $|a| = k$ is
\begin{equation}
\|u\|_{k,\Omega} = \left( \sum_{|a| \le k} \left\| \frac{\partial^{|a|} u}{\partial x_1^{a_1} \partial x_2^{a_2}} \right\|^2_{0,\Omega} \right)^{1/2}.
\end{equation}

\subsection{Finite element discretisation}

The finite element solution of the variational form for the scattering problem is sought in subsets $U_h, Z_h$ of the functional spaces
\begin{align}
U &= \left\{ u : u \in H^2(\Omega; \mathbb{C}) \text{ and } u = \nabla u \cdot \mathbf{n} = 0 \text{ on } \Gamma_g \right\},  
\\
\text{and}\quad
Z &= \left\{ (\zeta_1, \zeta_2) : \zeta_1 \in H^1(\Omega; \mathbb{C}), \zeta_2 \in H^1(V; \mathbb{C}) \text{ and } \zeta_1 = \zeta_2 \text{ on } \gamma \right\}. 
\end{align}
The discretised problem is to find $\eta^h \in U_h$ and $(\Phi^h, \varphi^h) \in Z_h$, such that
\begin{equation}
S_h([\chi^h, W^h, w^h], [\eta^h, \Phi^h, \varphi^h]) + i\omega \mathcal{A}_h([\chi^h, W^h], [\eta^h, \Phi^h]) = F_h(w^h), 
\end{equation}
for all $\chi^h \in U_h$ and $(W^h, w^h) \in Z_h$. 
The notations $S_h$, $\mathcal{A}_h$, and $F_h$ are used to account for possible nonconformity of the approximation. 
This includes the use of the nonconforming \citet{specht1988modified} approximation for the plate deflection in the sesquilinear form, that is
\begin{align}
a_h(\chi, \eta) &= \sum_{\kappa \in T_h} \int_{\Omega_\kappa} \nu K \Delta \bar{\chi} \, \Delta \eta \, dxdy \notag 
\\
&\quad + \sum_{\kappa \in T_h} \int_{\Omega_\kappa} K(1 - \nu)(\partial_{xx} \bar{\chi} \, \partial_{xx} \eta + \partial_{yy} \bar{\chi} \, \partial_{yy} \eta + 2\partial_{xy} \bar{\chi} \, \partial_{xy} \eta) dxdy. 
\end{align}
This formulation also applies to any inexact approximation of the computational domain or the integrals that appear in the functionals.

Two different types of finite element will be employed for the discretization of the scattering problem (Figure~\ref{fig:comp_domain}, right). 
In the ocean domain, $V$, a standard triangle with linear Lagrange shape functions and one degree of freedom per node (the potential $\varphi$) is used (LSWT). 
In the ice-shelf/cavity domain, $\Omega$, the hydroelastic element  \citep[SHEEL; introduced for ice-shelf modal analysis by][]{papathanasiou2019hydroelastic} is employed. SHEEL is a three-node triangle element, featuring four degrees of freedom per node: the ice shelf deflection, $\eta$, rotations, $\theta_x$ and $\theta_y$, and the velocity potential in the ice shelf cavity, $\Phi$. It uses linear Lagrange interpolation for $\Phi$, combined with the \citet{specht1988modified} nonconforming approximation for the plate deflection \citep{papathanasiou2019hydroelastic}.

The discretised finite element system has the form
\begin{equation}\label{eq:discrete_fem}
\begin{bmatrix}
\mathsfbi{K}_{\eta\eta} & -i\omega \mathsfbi{K}_{\eta\Phi} & \mathsfbi{0} \\
i\omega \mathsfbi{K}_{\Phi\eta} & \mathsfbi{K}_{\Phi\Phi} & \mathsfbi{K}_{\Phi\varphi} \\
\mathsfbi{0} & \mathsfbi{K}_{\varphi\Phi} & \mathsfbi{K}_{\varphi\varphi} - i\mathsfbi{K}_{RR}
\end{bmatrix}
\begin{bmatrix}
\boldsymbol{\eta} \\ \boldsymbol{\Phi} \\ \boldsymbol{\varphi}
\end{bmatrix}
=
\begin{bmatrix}
\boldsymbol{0} \\ \boldsymbol{0} \\ \boldsymbol{F}
\end{bmatrix}, 
\end{equation}
where the column vectors $\boldsymbol{\eta}$, $\boldsymbol{\Phi}$, and $\boldsymbol{\varphi}$ contain the nodal values for the ice shelf deflection and slope, $\eta$, $\theta_x$ and $\theta_{y}$ (all values at the grounding line excluded), the velocity potential in the cavity, $\Phi$, and the velocity potential in the ocean, $\varphi$, respectively. The matrices are such that $\mathsfbi{K}_{\Phi\eta} = \mathsfbi{K}_{\eta\Phi}^T$ and $\mathsfbi{K}_{\varphi\Phi} = \mathsfbi{K}_{\Phi\varphi}^T$, and the matrix $\mathsfbi{K}_{RR}$ is produced by the implementation of the DtN condition (or the Sommerfeld condition).

Assuming that $\omega^2 \mu < 1$, which is the case for the frequency range considered, there exists a constant $C > 0$, such that
\begin{equation}
a_h(\eta_h, \eta_h) + \int_\Omega g(1 - \omega^2 \mu) \bar{\eta}_h \eta_h \, d\Omega \geq C \left( \sum_{\kappa \in T_h} |\eta_h|_{2,\kappa}^2 + \|\eta_h\|_{0,\Omega}^2 \right). 
\end{equation}
This coercivity condition implies that $\mathsfbi{K}_{\eta\eta}$ is invertible \citep{rannacher1979nonconforming} and, therefore,
\begin{equation}\label{eq:eta_eqn}
\boldsymbol{\eta} = i\omega \mathsfbi{K}_{\eta\eta}^{-1} \mathsfbi{K}_{\eta\Phi} \boldsymbol{\Phi}. 
\end{equation}
Equation~\eqref{eq:eta_eqn} has implications for the deflection and velocity potential fields, which are
\begin{equation}\label{eq:re_im_eta}
\text{Re}[\boldsymbol{\eta}] = -\omega \mathsfbi{K}_{\eta\eta}^{-1} \mathsfbi{K}_{\eta\Phi} \, \text{Im}[\boldsymbol{\Phi}] \quad\text{and}\quad
\text{Im}[\boldsymbol{\eta}] = \omega \mathsfbi{K}_{\eta\eta}^{-1} \mathsfbi{K}_{\eta\Phi} \, \text{Re}[\boldsymbol{\Phi}]. 
\end{equation}
Equations~\eqref{eq:re_im_eta} indicate that, excluding values on the grounding line, the real part of the deflection is related to minus the imaginary part of the velocity potential in the ice shelf cavity and the imaginary part of the deflection is related to the real part of the velocity potential, respectively.

Setting $\mathsfbi{P}_{\Phi\Phi} = \mathsfbi{K}_{\eta\Phi}^T \mathsfbi{K}_{\eta\eta}^{-1} \mathsfbi{K}_{\eta\Phi}$ and using static condensation to eliminate $\boldsymbol{\eta}$ from the second block row  of system \eqref{eq:discrete_fem}, gives
\begin{equation}\label{eq:Ksys}
\begin{bmatrix}
\mathsfbi{K}_{\Phi\Phi} - \omega^2 \mathsfbi{P}_{\Phi\Phi} & \mathsfbi{K}_{\Phi\varphi} \\
\mathsfbi{K}_{\Phi\varphi}^T & \mathsfbi{K}_{\varphi\varphi} - i\mathsfbi{K}_{RR}
\end{bmatrix}
\begin{bmatrix}
\boldsymbol{\Phi} \\ \boldsymbol{\varphi}
\end{bmatrix}
=
\begin{bmatrix}
\boldsymbol{0} \\ \boldsymbol{F}
\end{bmatrix}. 
\end{equation}
This system is the scattering problem for the cavity with a correction term, $-\omega^2 \mathsfbi{P}_{\Phi\Phi}$, which adds the effect of the ice shelf flexure at the interior of $\Omega$. The statically condensed form \eqref{eq:Ksys} is an alternative method to solve system \eqref{eq:discrete_fem} but will not be pursued further in this study. However, it indicates that the system matrix produced in \eqref{eq:Ksys} is not Hermitian, which reflects the governing operator not being self-adjoint \citep{bennetts2021complex}.

\section{Model verification and parametric analysis}

Initial verification of the code developed for this study is made by removing the ice shelf and comparing results for the harbour problem with results in the existing literature (Appendix~\ref{app:harbour}).
Verification for the problem involving the ice shelf is made in \S\,\ref{sec:verification} by comparing results for the DtN and Sommerfeld conditions, by comparing the resonant peaks for a narrow ice shelf against results from a model with one horizontal dimension, and by using identity \eqref{eq:eta_eqn}.
This is followed by a parametric analysis on the aspect ratio of the ice shelf (\S\,\ref{sec:aspect_ratio}), the incident wave angle (\S\,\ref{sec:inc_angle}) and the grounding line geometry (\S\,\ref{sec:grounding_line}). 

Unless otherwise stated, the ice shelf thickness and bathymetry are constant, such that $H = 300\,\text{m}$ and $B = 900\,\text{m}$, and normally incident waves are used, $\alpha = 0$.
The ice shelf response to the incident wave forcing is shown in terms of the normalised flexural displacement, $(g/\omega)\,\eta(x,y)$, which corresponds to an incident displacement of unit-amplitude (in dimensions of the velocity potential, which are neglected in the presentation for the sake of brevity).
Hydroelastic response spectra, $(g/\omega)\|\eta\|_{\infty}$ where $\|\eta\|_{\infty} = \max_{\Omega} |\eta|$ (i.e., the maximum amplification of the ice shelf with respect to the incident amplitude), are shown as functions of the nondimensional wavenumber, $kL$, where $L$ is a characteristic ice shelf length.
(Dimensional wave periods are indicated for the resonant responses.) 
The nondimensional wavenumber ranges considered correspond to wave periods from approximately 0.1\,h to 3\,h, i.e., the long-period wave regime (Figure~\ref{fig:fig1}, bottom).

\subsection{Preliminary results and verification}\label{sec:verification}

\begin{figure}
    \centering
    \setlength{\figurewidth}{0.85\textwidth} 
    \setlength{\figureheight}{0.4\textwidth} 
    \input{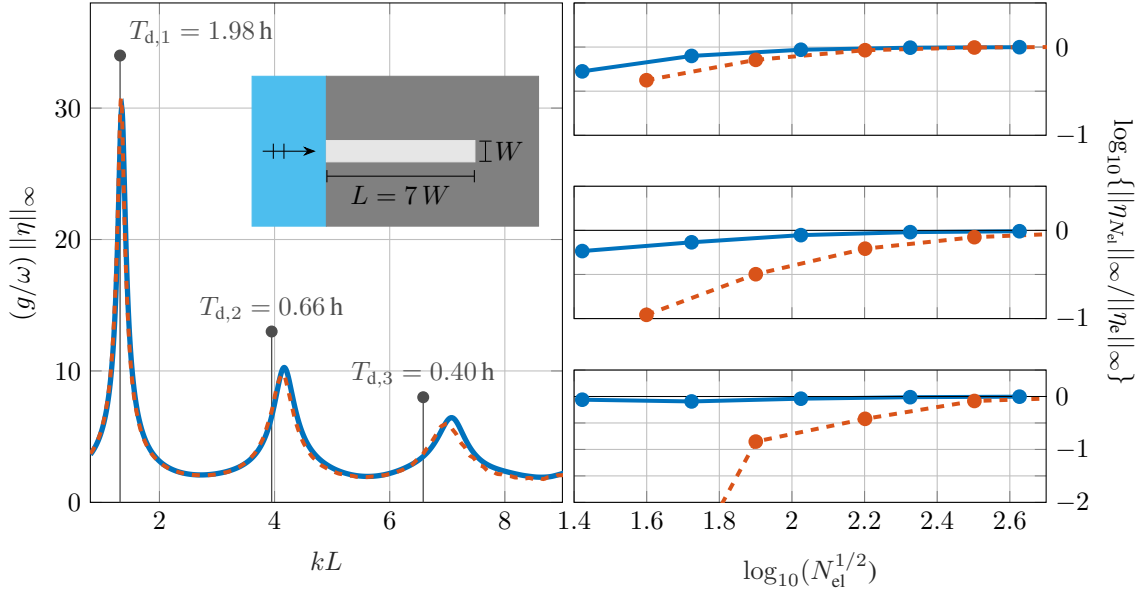} 
    \caption{(Left)~Hydroelastic spectrum, i.e., ice shelf flexure amplitude as a function of the nondimensional wave number, using DtN condition ($R=2\,L$, $n=10$, $N_{\text{el}}=11,200$; solid blue curve) and Sommerfeld condition ($R=20\,L$, $N_{\text{el}}=404,480$; broken red curve), with 1D model resonant frequencies, $T_{\textrm{d},j}$ for $j=1,2,3$ \eqref{eq:6.1}, given for comparison (black bullets on stems). 
    Inset shows the geometry considered, with a rectangular ice shelf (light grey; $L=140$\,km and $W=20$\,km) of thickness $H = 300\,\text{m}$  embedded into the grounded ice sheet (dark grey), forced by normally incident waves from the open ocean (blue), with constant bathymetry $B = 900\,\text{m}$.
    (Right)~Convergence of ice shelf amplitude as a function of the number of finite elements used, for DtN condition (solid blue) and Sommerfeld condition (broken red), at wave periods $T_{\textrm{d},j}$ for $j=1$ (top), $j=2$ (middle) and $j=3$ (bottom).}
    \label{fig:fig4}
\end{figure}

Consider a narrow, rectangular ice shelf of length $L = 140\,\text{km}$ and width $W = 20\,\text{km}$. 
The ice shelf is fully embedded, such that the grounding line spans three of its four sides, including the two sides of length $L$ (Figure~\ref{fig:fig4}). 
Both Sommerfeld and DtN options are tested. 
In the case of the Sommerfeld condition, $R = 20L$ is selected, which is large enough to give convergent results for the frequency band tested. 
For DtN, $R = 2L$ with $n = 10$ series terms, where convergence is verified via numerical experiments  with $n = 5$--$100$ (not shown).

Figure~\ref{fig:fig4} (left) shows the hydroelastic spectrum for the rectangular ice shelf. 
Solutions using the Sommerfeld and DtN conditions agree. 
Three resonant peaks of increased amplitude occur over the interval shown. The first resonant peak occurs very close to $T_{\textrm{d},1} = 1.98$\,h, where $T_{\textrm{d},n}$ are the harmonics for a model with one horizontal dimension and a Dirichlet boundary condition in the water column below the shelf front, which are \citep{papathanasiou2019hydroelastic}
\begin{equation}
T_{\textrm{d},n} = \frac{L}{1800\sqrt{gB(n - 1/2)^2 \left(1 - \frac{\rho_{\textrm{i}}}{\rho_{\textrm{w}}} \frac{H}{B} \right)}} \quad \text{for} \quad n = 1, 2, 3, \ldots
\label{eq:6.1}
\end{equation}
The second resonant peak is reasonably well approximated by \eqref{eq:6.1} at $T_{\textrm{d},2} = 0.66$\,h, while the shift is significant for the third resonance, where the prediction of $T_{\textrm{d},3} = 0.40$\,h from \eqref{eq:6.1} is below the peak at $kL \approx 7$. 
This is attributed to multidimensional effects that become more important as frequency increases and modes of deflection along the ice shelf width contribute more to the overall response.

Figure~\ref{fig:fig4} (right) demonstrates convergence of the numerical solution for different quasi-uniform mesh sizes, $\log_{10}(\sqrt{N_{\text{el}}})$, where $N_{\text{el}}$ is the number of elements in each mesh. Convergence is established by $\log_{10}(\|\eta_{N_{\text{el}}}\|_{\infty}/\|\eta_{\textrm{e}}\|_{\infty}) \to 0$ with increasing $N_{\text{el}}$, where $\eta_{\textrm{e}}$ is a solution using a fine mesh ($N_{\text{el}} = 404{,}480$ with the Sommerfeld condition). 
In all cases, the DtN condition is more efficient than the Sommerfeld. In the large-$N_{\text{el}}$ regime, the error increases with increasing frequency, as expected, considering that more nodes are needed per wavelength to represent the wave field accurately.

\begin{figure}
    \centerline{
     \includegraphics[width=0.99\linewidth]{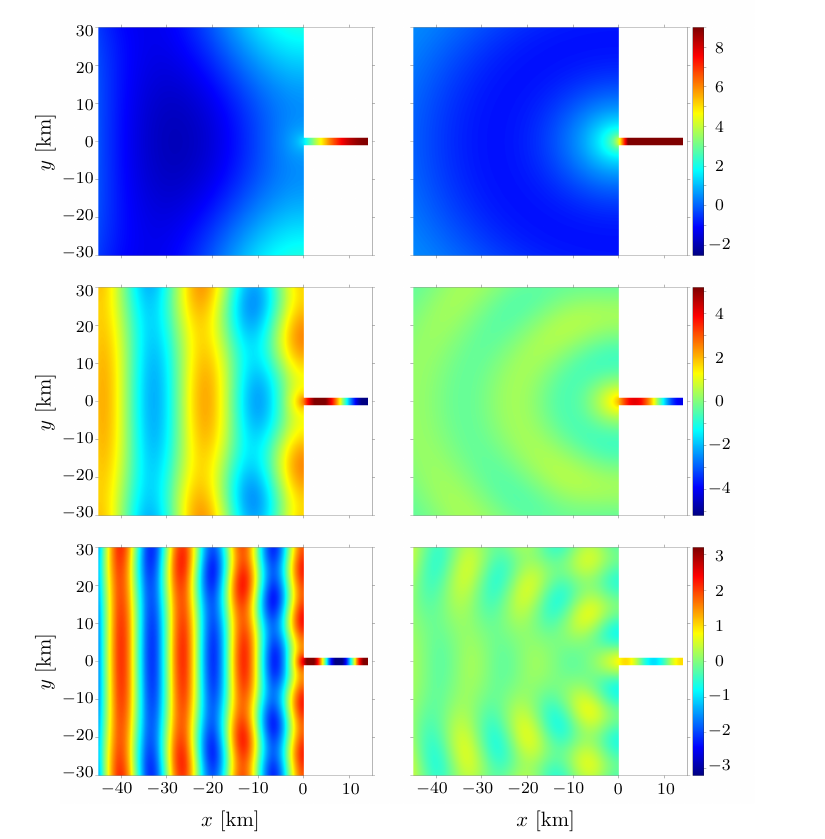}
    }
    \caption{Real parts (left) and imaginary parts (right) of velocity potential in the ice shelf/cavity region ($\Phi(x,y)$, $x>0$) and surrounding open water ($\varphi(x,y)$, $x<0$), for the problem in Fig.~\ref{fig:fig4} at wave periods $T = T_{\textrm{d},1} = 1.98$\,h (first row; resonant period), $T_{\textrm{d},2} = 0.66$\,h (second row; close to resonance), and $T_{\textrm{d},3} = 0.40$\,h (third row; non-resonant). The colorbar for the imaginary part of the resonant case (top-right panel) does not extend to the maximum value of the velocity potential in the ice shelf/cavity region, $\textrm{Im}(\Phi)\approx 30$. }
    \label{fig:fig5}
\end{figure}

Figure~\ref{fig:fig5} shows the 2D field of the velocity potential in the ice shelf/cavity region  and the immediate surrounding open ocean. 
Results are shown for $T = T_{\textrm{d},1} = 1.98$\,h (the first resonant peak; first row), $T_{\textrm{d},2} = 0.66$\,h (close to second resonant peak; second row), and $T_{\textrm{d},3} = 0.40$\,h (non-resonant; third row). 
The first column in each case corresponds to the real part of the response and the second column to the imaginary part. 
For the resonant period, the potential is significantly amplified in the ice shelf/cavity region  ($\textrm{Im}(\Phi)>8$ generally)  but reduces significantly for the other two periods. 
Similarly, the deflection is amplified at the resonant period (not shown), noting that formula \eqref{eq:eta_eqn} shows within the ice shelf horizontal span (not in the vicinity of the grounding line), the real part of the deflection follows the imaginary part of the velocity potential, and the imaginary part of the deflection follows the real part of the velocity potential but with the opposite sign.

\subsection{Ice shelf aspect ratio}\label{sec:aspect_ratio}

\begin{figure}
    \centering
    \setlength{\figurewidth}{0.9\textwidth} 
    \setlength{\figureheight}{0.5\textwidth} 
    \input{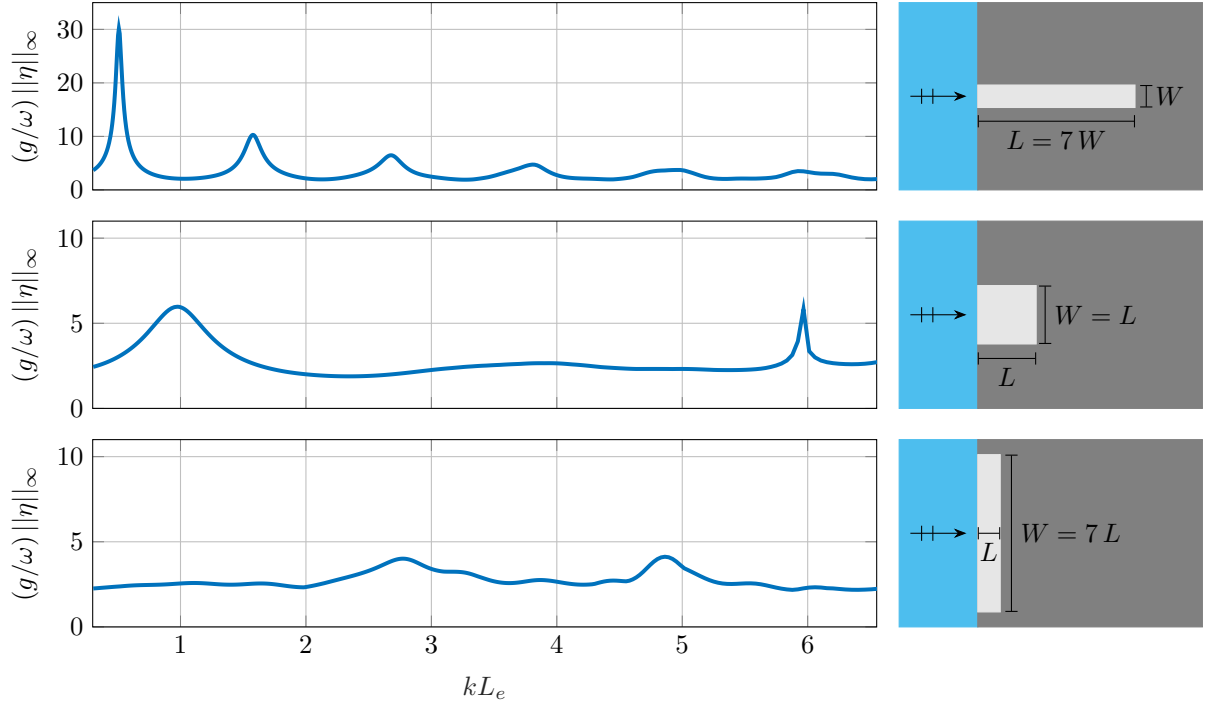} 
    \caption{(Left)~Hydroelastic spectra forced by normally incident waves ($\alpha=0$) for ice shelves of the same area ($A = 2800\,\text{km}^2$) and thickness ($H=300\,\text{m}$) and over the same constant bathymetry ($B=900\,\text{m}$), but where the ice shelves have different shapes, as shown by corresponding geometries (right), with $W=20\,\text{km}$ (top), $W\approx{}53\,\text{km}$ (middle) and $W=140\,\text{km}$ (bottom).}
    \label{fig:fig6}
\end{figure}

Three different fully embedded ice shelves are studied: a narrow rectangular ice shelf, a square ice shelf, and a wide rectangular ice shelf (Figure~\ref{fig:fig6}, right). 
The ice shelves share the same horizontal span of $A = 2800\,\text{km}^2$. The dimensions of the narrow ice shelf are $L = 140\,\text{km}$ and $W = 20\,\text{km}$. The side of the square ice shelf is $L = W = \sqrt{A} \equiv L_{\textrm{e}} \approx 53~\text{km}$. The wide ice shelf has $L = 20~\text{km}$ and $W = 140~\text{km}$. 

Hydroelastic spectra are shown as functions of $kL_{\textrm{e}}$.
The results 
show narrow and elongated ice shelves have more distinct resonant peaks, and the fundamental mode features the maximum amplitude overall. 
As the ice shelf front becomes wider, the peaks become less pronounced. 
Further, the maximum amplitude reduces from the narrow to the square to the wide ice shelf, whilst the response spectrum curve becomes flatter.
The phenomenon of narrower embedded ice shelves experiencing larger responses is related to the harbour paradox, in which narrower harbour opening can create larger amplitude responses within the harbour \citep[][Chap~5]{mei2005theory}.

\subsection{Incidence angle}\label{sec:inc_angle}

\begin{figure}
    \centerline{
    \includegraphics[width=0.99\linewidth]{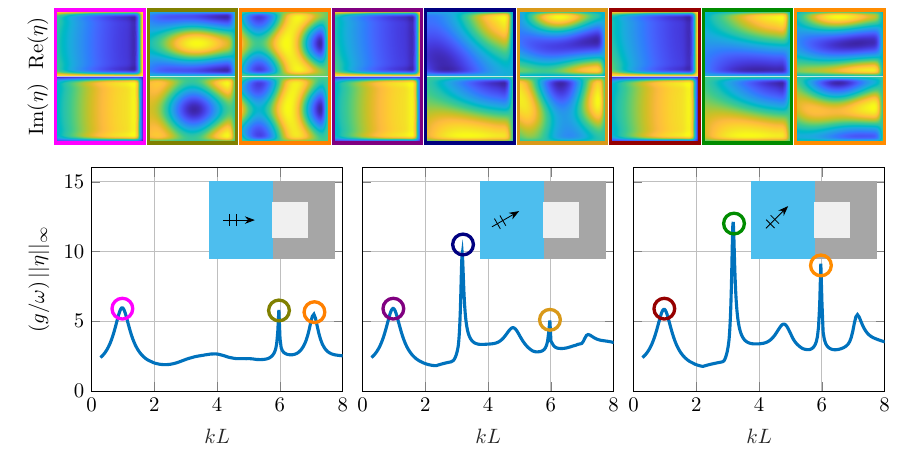}
    }
    \caption{(Bottom)~Hydroelastic spectra for a square ice shelf of side length $L = 53~\text{km}$ and thickness $H = 300~\text{m}$ over constant bathymetry $B = 900~\text{m}$, forced by incident waves at angles (left)~$\alpha=0$, (middle)~$\alpha=\pi/6$ and (right)~$\alpha=\pi/4$ (as shown by insets). 
    (Top)~Real and imaginary parts of ice shelf deflection for resonances, where the colours of the bounding boxes correspond to the colours of the circles on the hydroelastic spectra, indicating the relevant resonances.}
    \label{fig:fig7}
\end{figure}

Figure~\ref{fig:fig7} presents hydroelastic spectra for a square shelf (side length $L = 53~\text{km}$), as a function of the nondimensional wavenumber, $kL$, for three different angles of incidence, $\alpha = 0$, $\pi/6$ and $\pi/4$. 
Contour plots of the real and imaginary parts of $\eta(x,y)$ at several resonant peaks of the spectrum are included. 
The angle of incidence has virtually no effect on the fundamental (lowest frequency) mode, but the remaining part of the spectrum is alters significantly in response to changes in the incidence angle. 
In particular, the response at some higher harmonics is enhanced for oblique incidence, e.g., the peak at $kL \approx 3$ is amplified as the angle increases to $\pi/4$, while the deflection pattern is different at the common peak occurring at $kL \approx 6$ when $\alpha$ changes from $0$ to $\pi/6$.

\begin{figure}
    \centering
    \setlength{\figurewidth}{0.95\textwidth} 
    \setlength{\figureheight}{0.5\textwidth} 
    \input{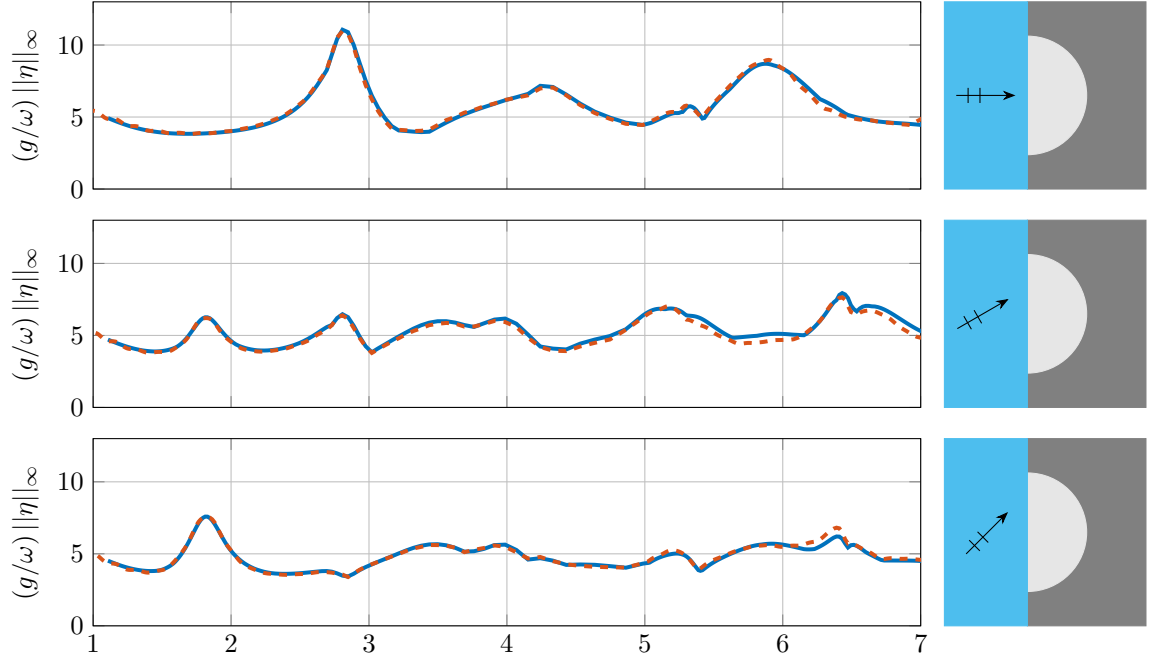} 
    \caption{Hydroelastic spectra of a semi-circular ice shelf of radius $L = 200~\text{km}$ and thickness $H = 300~\text{m}$ over constant bathymetry $B = 500~\text{m}$ (similar to the Larsen~C Ice Shelf), forced by incident waves at angles (top)~\(\alpha = 0\), (middle)~\(\alpha=\pi/6\), and (bottom)~\(\alpha=\pi/4 \).
    Results are given using the DtN condition (solid blue curves) and Sommerfeld condition (broken red curves).}
    \label{fig:fig9}
\end{figure}

The effect of incidence angle is also studied for a semi-circular ice shelf of radius \( L = 200 \, \text{km} \), thickness \( H = 300 \, \text{m} \), and bathymetry \( B = 500 \, \text{m} \) (Figure~\ref{fig:fig9}), 
which resembles the Larsen~C Ice Shelf. 
Hydroelastic spectra are shown for incidence angles \( \alpha = 0, \pi/6, \pi/4 \). 
(The Sommerfeld and DtN conditions are checked for this semi-circular problem and are found to be in good agreement.) 
The angle of incidence is again found to change the form of the amplitude spectrum and, in particular, the location of the resonant peaks. For example, when \( \alpha = 0 \), at \( kL \approx 2.8 \) (period \( T = 1.78 \)\,h) the first resonant peak appears, where
\[
(g/\omega)\, \| \eta \|_{\infty}\approx 10.
\]
The same period corresponds to the second resonant peak when \( \alpha = \pi/6 \) and the amplitude is reduced to
\[
(g/\omega)\, \| \eta \|_{\infty} \approx 5.
\]
There is no resonant peak at this period when \( \alpha = \pi/4 \).

\subsection{Grounding line geometry: Ice shelves vs ice tongues}\label{sec:grounding_line}

\begin{figure}
    \centerline{
    \begin{tabular}{cc}
    \includegraphics[width=0.25\linewidth]{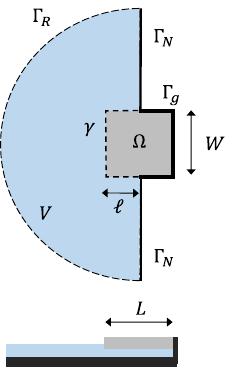} 
    &
    \\[-160pt]
    &
    \includegraphics[width=0.475\linewidth]{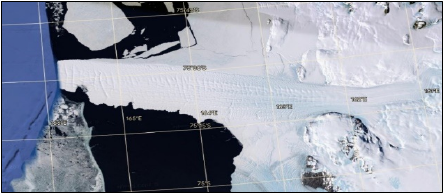}
    \\
    & 
    \includegraphics[width=0.475\linewidth]{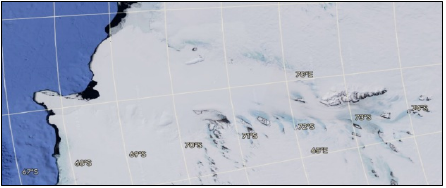}
    \end{tabular}
    }
    \caption{(Left)~Computational domain for the scattering problem of rectangular formations with different grounding line extent. (Right)~Drygalski Ice Tongue (top) and Amery Ice Shelf (bottom).}
    \label{fig:fig11}
\end{figure}

Figure~\ref{fig:fig11} (right) depicts two formations in Antarctica, namely the Drygalski Ice Tongue (upper image) and Amery Ice Shelf (lower image).
In the case of the Drygalski Ice Tongue, the grounding line (highlighted with a black line) 
contains only around 30\% of the ice shelf length.
In contrast, Amery Ice Shelf is fully embedded. 
This observation motivates a study of the influence of the proportion of the ice shelf protrusion.
Rectangular ice shelves are studied, of length $L=140$\,km and width $W=20$\,km (as in \S~\ref{sec:verification}), where segments of length $0 \leq l \leq L$ protrude into the open ocean (Figure~\ref{fig:fig11}, left). 
(The ice thickness and bathymetry are $H = 300$\,m. and $B = 900$\,m , respectively.)
 
\begin{figure}
    \centering
    \setlength{\figurewidth}{0.85\textwidth} 
    \setlength{\figureheight}{0.5\textwidth} 
    \input{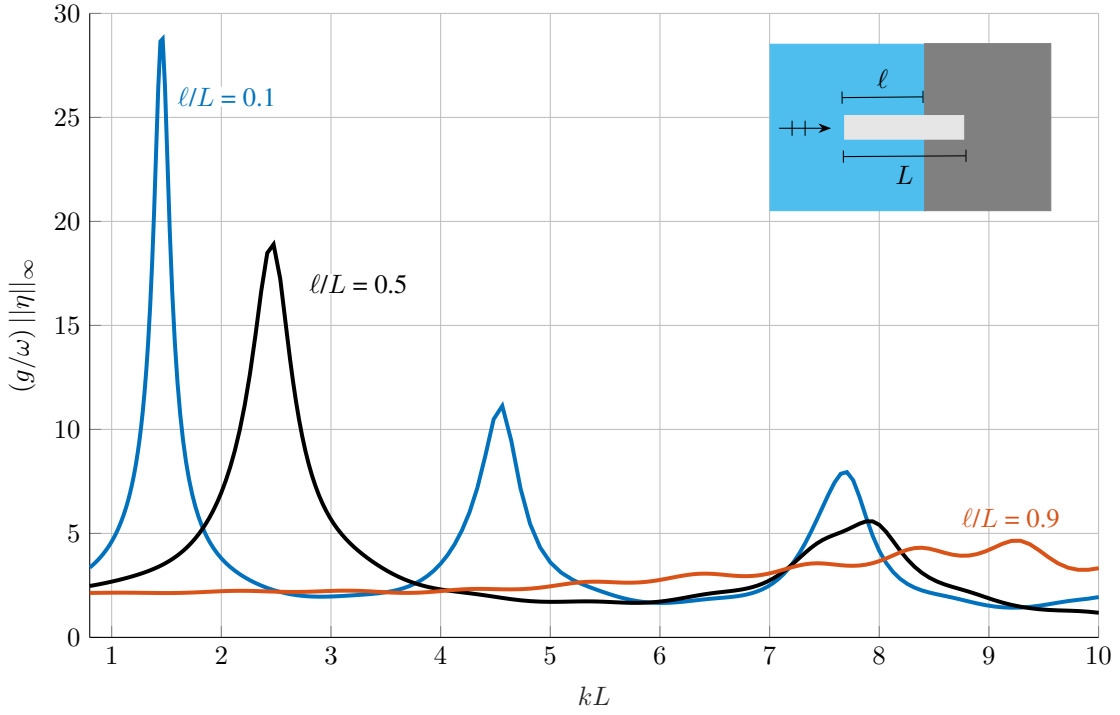} 
    \caption{Hydroelastic spectra of narrow rectangular ice shelves of length $L$, with constant thickness ($H=300\,\text{m}$), over constant bathymetry ($B=900\,\text{m}$) and in response to normally incident waves ($\alpha = 0$). The protruding proportions are  $l/L=0.1$ (blue), 0.5 (black), and 0.9 (red). 
}
    \label{fig:fig12}
\end{figure}

Figure~\ref{fig:fig12} presents hydroelastic spectra forced by normally incident waves, with $l/L$ values 0.1, 0.5, and 0.9, ranging from an almost fully embedded ice shelf to a mostly protruding ice tongue.  
As the protruding proportion, $l/L$, increases, the resonant frequency peaks decrease in amplitude and shift towards higher frequencies (larger $kL$ values).

\begin{figure}
    \centerline{
    \includegraphics[width=0.99\linewidth]{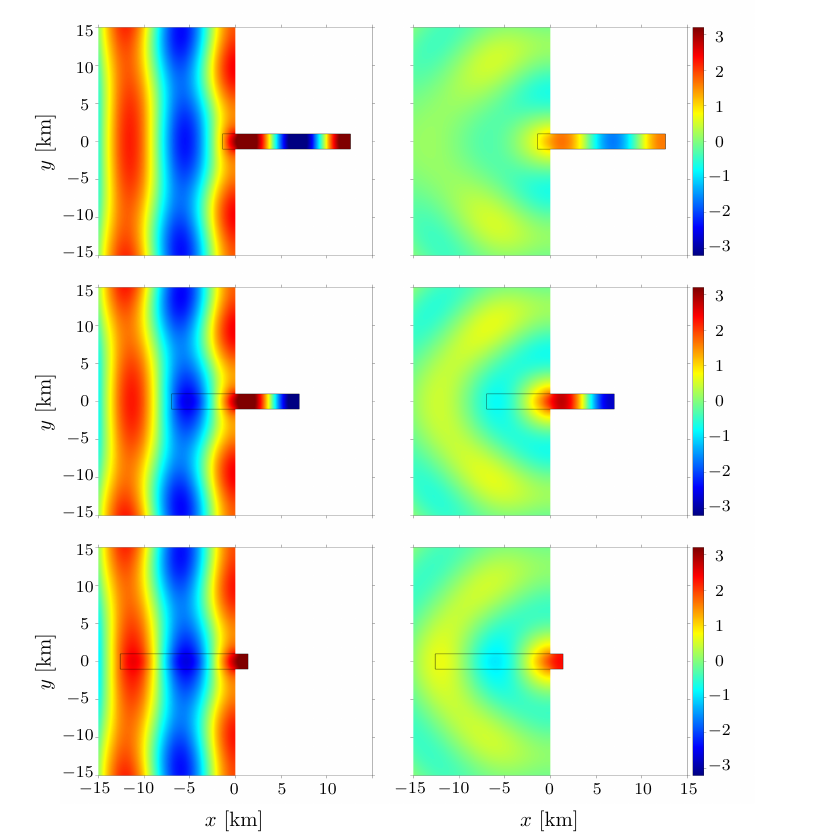}
    }
    \caption{Similar to Figure~\ref{fig:fig5}, but for ice shelves in Fig.~\ref{fig:fig12}, i.e., where the ice shelf protrudes from the coastline, with (top)~$l/L=0.1$, (middle)~$l/L=0.5$, and (bottom)~$l/L=0.9$, at $T=0.35$\,h.
    The perimeters of the ice shelves are indicated with black lines.}
    \label{fig:fig13}
\end{figure}

Figure~\ref{fig:fig13} displays the velocity potential field for the three cases at wave period $T = 0.35$\,h
($kL\approx 7.43$, which is close to the highest frequency resonances for $\ell/L=0.1$ and 0.5; Fig.~\ref{fig:fig12}). 
The locations of the deflection peaks and troughs change for different values of $l/L$. 
The response of the unconfined part of the shelf has practically the same profile as the open ocean surface gravity waves. This is due to the slenderness of the chosen rectangle.  
Larger responses tend to occur in the region where the ice shelf is embedded.

\section{Conclusions}

A solution method has been presented for a water wave scattering problem involving the hydroelastic response of an ice shelf embedded along a flat coastline to incident waves from the open ocean. The proposed approach combines a finite element method, where nonconforming hydroelastic finite element triangles are used for the high-order ice shelf–water cavity region, and a Dirichlet-to-Neumann mapping to apply the radiation condition in the open ocean. The solution is efficient enough to study the hydroelastic response spectra of the ice shelf to incident wave forcing over the broad wavenumber range.
Numerical results were used to conduct a novel study on how the ice shelf response is influenced by the incident wave angle, the ice shelf shape and the proportion of the ice shelf boundary attached to land. 
In all cases, it was shown that significant effects in the response spectra could be produced. 
The findings indicate the potential importance of models that include non-uniform motions in the span-wise direction. 

The numerical method presented could be extended to accommodate a model in which the coastline surrounding the ice shelf is not straight, which may influence ice shelf flexure, e.g., in cases when it is in a bay (such as the Amery Ice Shelf, Fig.~\ref{fig:fig1}a). 
In particular, the reflected wave field (Eq.~\ref{eq:varphi_decomp}) would need to be modified, as in \citet{steward2001improved}.
Similarly, the varying bathymetry could be extended beyond the ice shelf/cavity region, although this would require modifications of the DtN condition.
The method could also be extended to model partially absorbing coastlines, i.e., changing from a Neumann to a Robin boundary condition \citep{isaacson1990waves}.

The shallow-water approximation (\ref{eq:plate3_freq}) is unlikely to be accurate for the incident ocean swell that have been heavily implicated in major ice shelf calving events \citep{massom2018disintegration,teder2025large}, as swell often have wave periods $<15$\,s.  
Replacing the shallow-water approximation with the so-called single-mode approximation \citep{bennetts2007multi} would extend the validity of the model into the swell regime \citep{bennetts2021complex} and maintain the shallow-water structure of the governing equations, such that the solution method could be adapted to it in a straightforward manner.
However, incorporating the bending moments applied by incident swell on shelf fronts \citep{bennetts2024thin,bennetts2026modelling} in the model would require modifications to the method.
Further, \citet{chen2018ocean} observed that flexural waves forced by swell attenuate with distance travelled into ice shelves. 
This implies viscosity in the ice shelf response should be considered in the swell regime, which could be incorporated into the model used in the presented study using, e.g., the 
viscoelastic model for ice shelf flexure proposed by \citet{macayeal2015model}.

In conclusion, we have contributed a solution method for a model of hydroelastic flexure of ice shelves excited by long ocean waves, where variations in both horizontal dimensions are incorporated and the ice shelf is of arbitrary shape, including varying thickness.
The method is efficient enough to study the ice shelf response spectra and identify resonances. 
Moreover, we have contributed new understanding of the dependence of the ice shelf response to incident waves on parameters that exist due to the second spatial dimension.
Our study sets the computational basis for understanding which Antarctic ice shelves are most susceptible to impacts by long ocean waves.

\section*{Acknowledgements}

The authors thank the Isaac Newton Institute for Mathematical Sciences, Cambridge, for support and hospitality during the program Multiple Wave Scattering supported by EPSRC grant no EP/R014604/1, where this work was initiated. LGB and MHM are funded by the Australian Research Council (FT190100404, SR200100008, DP240100325, DP240102104).

\section*{Data availability}

The data for the study is freely available, see \cite{papathanasiou2025modelling}.

\appendix

\section{Neglect of Coriolis effects}\label{append:coriolis}

Assuming an average southern latitude of $80^\circ$, the Coriolis frequency is given by
\begin{equation}
f = 2\Omega_e \sin(80^\circ),
\end{equation}
where $\Omega_e$ is the angular velocity of Earth's rotation. This implies a Coriolis period of approximately
\begin{equation}
T_f \approx \SI{12}{\hour}.
\end{equation}
The maximum wave period in the band of interest is $T_m \approx \SI{4}{\hour}$, and the corresponding Rossby number is
\begin{equation}
\mathrm{Ro} = \frac{T_f}{T_m} \approx 3.
\end{equation}
This Rossby number indicates that the Coriolis force effect is weak compared to other restoring forces (such as gravity and acceleration) in the wave-band of interest (15\,s to 4\,h). 
Therefore, Coriolis effects are not included in the study.

\section{Derivation and implementation of the DtN condition}\label{append:DtN}

The incident and reflected wave, using the Jacobi-Anger expansion, can be written as
\begin{equation}
\phi_I = e^{ikr(\cos\theta \cos\alpha + \sin\theta \sin\alpha)} = e^{ikr\cos(\alpha - \theta)} = \sum_{n=-\infty}^{\infty} i^n J_n(kr) e^{in(\alpha - \theta)},
\end{equation}
and
\begin{equation}
\phi_R = e^{ikr(-\cos\theta \cos\alpha + \sin\theta \sin\alpha)} = e^{-ikr\cos(\alpha + \theta)} = \sum_{n=-\infty}^{\infty} (-1)^n i^n J_n(kr) e^{in(\alpha + \theta)},
\end{equation}
where \( J_n(kr) \) is the Bessel function of the first kind and order \( n \). Then,
\begin{equation}
\phi_I + \phi_R = \sum_{n=-\infty}^{\infty} i^n e^{in\alpha} J_n(kr) \Lambda_n(\theta),
\end{equation}
where
\begin{equation}
\Lambda_n(\theta) = 
\begin{cases}
2 \cos(n\theta), & \text{for } n \text{ even}, \\
-2i \sin(n\theta), & \text{for } n \text{ odd}.
\end{cases}
\end{equation}
Since \( \partial \phi / \partial \theta = 0 \) at \( \theta = \pi/2, 3\pi/2 \), and using \( J_{-n}(kr) = (-1)^n J_n(kr) \), it follows
\begin{equation}
\phi_I + \phi_R = \sum_{n=0,2,4,\dots}^{\infty} A_n \frac{J_n(kr)}{J_n(kR)} \cos(n\theta) + \sum_{n=1,3,5,\dots}^{\infty} B_n \frac{J_n(kr)}{J_n(kR)} \sin(n\theta),
\end{equation}
with
\begin{subequations}
\begin{equation}
A_0 = 2 J_0(kR), \quad
A_n = 4 i^n \cos(n\alpha) J_n(kR), \quad n = 2, 4, 6, \dots
\end{equation}
\begin{equation}
\text{and}\quad
B_n = 4 i^n \sin(n\alpha) J_n(kR), \quad n = 1, 3, 5, \dots.
\end{equation}
\end{subequations}
The scattering potential \( \phi_s \) is now expanded as:
\begin{equation}
\phi_s = \sum_{n=0,2,4,\dots}^{\infty} C_n \frac{H_n^{(1)}(kr)}{H_n^{(1)}(kR)} \cos(n\theta) + \sum_{n=1,3,5,\dots}^{\infty} D_n \frac{H_n^{(1)}(kr)}{H_n^{(1)}(kR)} \sin(n\theta),
\end{equation}
where \( H_n^{(1)}(kr) \) is the Hankel function of the first kind and order \( n \). At \( r = R \), we have
\begin{equation}
\phi(R, \theta) = \phi_I(\theta) + \phi_R(\theta) + \phi_s(\theta) = \sum_{n=0,2,4,\dots}^{\infty} (A_n + C_n) \cos(n\theta) + \sum_{n=1,3,5,\dots}^{\infty} (B_n + D_n) \sin(n\theta).
\end{equation}
Using the orthogonality conditions for trigonometric functions, the coefficients are
\begin{subequations}
\begin{equation}
A_0 + C_0 = \frac{1}{\pi} \int_{\pi/2}^{3\pi/2} \phi(\vartheta) d\vartheta = I_0,
\end{equation}
\begin{equation}
A_n + C_n = \frac{2}{\pi} \int_{\pi/2}^{3\pi/2} \cos(n\vartheta) \phi(\vartheta) d\vartheta = I_n, \quad n = 2, 4, 6, \dots
\end{equation}
\begin{equation}
\text{and}\quad
B_n + D_n = \frac{2}{\pi} \int_{\pi/2}^{3\pi/2} \sin(n\vartheta) \phi(\vartheta) d\vartheta = K_n, \quad n = 1, 3, 5, \dots.
\end{equation}
\end{subequations}
Then, at \( r = R \), the radial derivative is
\begin{subequations}
\begin{align}
\left. \frac{\partial \phi}{\partial r} \right|_{r=R}  & =
\sum_{n=0,2,4,\dots}^{\infty} \left[ \frac{A_n W_n}{J_n(kR) H_n^{(1)}(kR)} + I_n \frac{H_n^{(1)'}(kR)}{H_n^{(1)}(kR)} \right] \cos(n\theta)
\\
& ~
+ \sum_{n=1,3,5,\dots}^{\infty} \left[ \frac{B_n W_n}{J_n(kR) H_n^{(1)}(kR)} + K_n \frac{H_n^{(1)'}(kR)}{H_n^{(1)}(kR)} \right] \sin(n\theta)
\end{align}
\end{subequations}
The Wronskian \( W_n \) is
\begin{equation}
W_n = J_n'(kR) H_n^{(1)}(kR) - H_n^{(1)'}(kR) J_n(kR) = -\frac{2i}{\pi R}.
\end{equation}
The normal derivative at \( r = R \) then has the form
\begin{equation}
\left. \frac{\partial \phi}{\partial r} \right|_{r=R} = f(R, \theta) + \int_{\pi/2}^{3\pi/2} G(R, \theta, \vartheta) \phi(R, \vartheta) d\vartheta,
\end{equation}
where the function \( f(R, \theta) \) is
\begin{equation}
f(R, \theta) = \sum_{n=0,2,4,\dots}^{\infty} \frac{A_n W_n \cos(n\theta)}{J_n(kR) H_n^{(1)}(kR)} + \sum_{n=1,3,5,\dots}^{\infty} \frac{B_n W_n \sin(n\theta)}{J_n(kR) H_n^{(1)}(kR)},
\end{equation}
and the kernel function is
\begin{equation}
G(R, \theta, \vartheta) = \frac{M_0}{\pi} + \frac{2}{\pi} \sum_{n=2,4,6,\dots}^{\infty} M_n \cos(n\theta) \cos(n\vartheta) + \frac{2}{\pi} \sum_{n=1,3,5,\dots}^{\infty} M_n \sin(n\theta) \sin(n\vartheta),
\end{equation}
with
\begin{equation}
M_n = \frac{H_n^{(1)'}(kR)}{H_n^{(1)}(kR)} = \frac{n}{R} - k \frac{H_{n+1}^{(1)}(kR)}{H_n^{(1)}(kR)}.
\end{equation}

When the DtN condition is applied on \( \Gamma_R \), the following integrals need to be computed
\begin{equation}
I_1 = \int_{\Gamma_R} B \bar{w}^h \left( \int_{\pi/2}^{3\pi/2} G(R, \theta, \vartheta) \phi(R, \vartheta) \, d\vartheta \right) d\Gamma, \quad\text{and}\quad
I_2 = \int_{\Gamma_R} B \bar{w}^h f(R, \theta) \, d\Gamma.
\end{equation}
These integrals are computed numerically in the MATLAB code used in this study. The accuracy of the numerical integration matches the accuracy of the finite element formulation, so that the overall convergence rate is not affected. In the following, \( \Gamma_R \) is assumed to be composed of \( N \) element edges for elements \( e_j \), \( j = 1, 2, \dots, N \). Starting with the second integral and denoting each term of the series expansion of \( f \) as \( f_n \), it is
\begin{equation}
I_2 \approx \sum_{n=0}^{\infty} \sum_{j=1}^{N} \int_{e_j} B \bar{w} f_n(R, \theta) \, d\Gamma.
\end{equation}
For each element on \( \Gamma_R \), the forcing vector will have two nonzero terms, which are approximated using the midpoint rule and the resulting linear Lagrange shape functions on the element edge (\( \theta_j = \arccos(x_j / R) \) is the angle corresponding to the midpoint of the \( e_j \) element edge, and \( h_j \) is the element edge length) as
\begin{equation}
f_1 = f_2 \approx \frac{B}{2} h_j \sum_{n=0}^{\infty} f_n(R, \theta_j).
\end{equation}
For the first integral \( I_1 \), the inner part is considered first. It is
\begin{equation}
\int_{\pi/2}^{3\pi/2} G(R, \theta, \vartheta) \phi(R, \vartheta) \, d\vartheta = 
\sum_{n=0}^{\infty} \sum_{i=1}^{N} \int_{e_i} G_n(R, \theta, \vartheta) \phi(R, \vartheta) \, d\vartheta
\approx \frac{1}{R} \sum_{n=0}^{\infty} \sum_{i=1}^{N} G_n(R, \theta, \vartheta_i) \phi(R, \vartheta_i) h_i.
\end{equation}
However, the value \( \phi(R, \vartheta_i) \) is not a nodal degree of freedom. At the same order of approximation with the elements used,
\begin{equation}
\phi(R, \vartheta_i) = \frac{\phi(R, \vartheta_{A_{1i}}) + \phi(R, \vartheta_{A_{2i}})}{2} = \frac{\phi_{A_{1i}} + \phi_{A_{2i}}}{2}
\end{equation}
for an element with nodes on \( \Gamma_R \) denoted \( A_1, A_2 \). Then, handling the outer integral as in the case of \( I_2 \), the local element contributions from the application of the DtN condition to the stiffness matrix and force vector are (for element \( e_j \) on \( \Gamma_R \)):
\begin{equation}
I_1 \big|_{e_j} \approx \frac{B}{4R} \sum_{i=1}^{N} \sum_{n=0}^{\infty} G_n(R, \theta_j, \vartheta_i) h_i h_j
\begin{bmatrix}
1 & 1 \\
1 & 1
\end{bmatrix}
\begin{bmatrix}
\phi_{A_{1i}} \\
\phi_{A_{2i}}
\end{bmatrix},
\end{equation}
\begin{equation}
I_2 \big|_{e_j} \approx \frac{B}{2} \sum_{n=0}^{\infty} f_n(R, \theta_j) h_j
\begin{bmatrix}
1 \\
1
\end{bmatrix}.
\end{equation}
The implementation is done by truncating the infinite series over \( n \) to a term \( N_n \) and assembling the element contributions over \( \Gamma_R \) to the global stiffness matrix and force vector respectively.

\section{Validation for the harbour problem}\label{app:harbour}

\begin{figure}
    \centerline{
    \begin{tabular}{c @{\hspace{15pt}} c}
    \includegraphics[height=7cm]{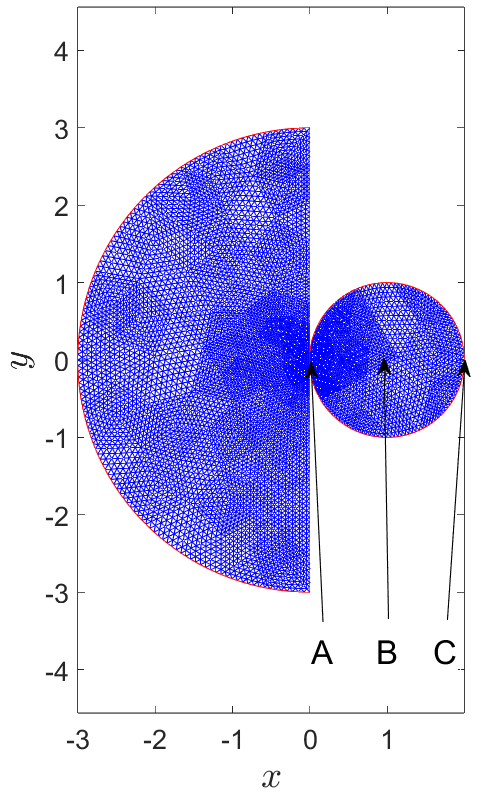}
    &
    \includegraphics[height=7cm]{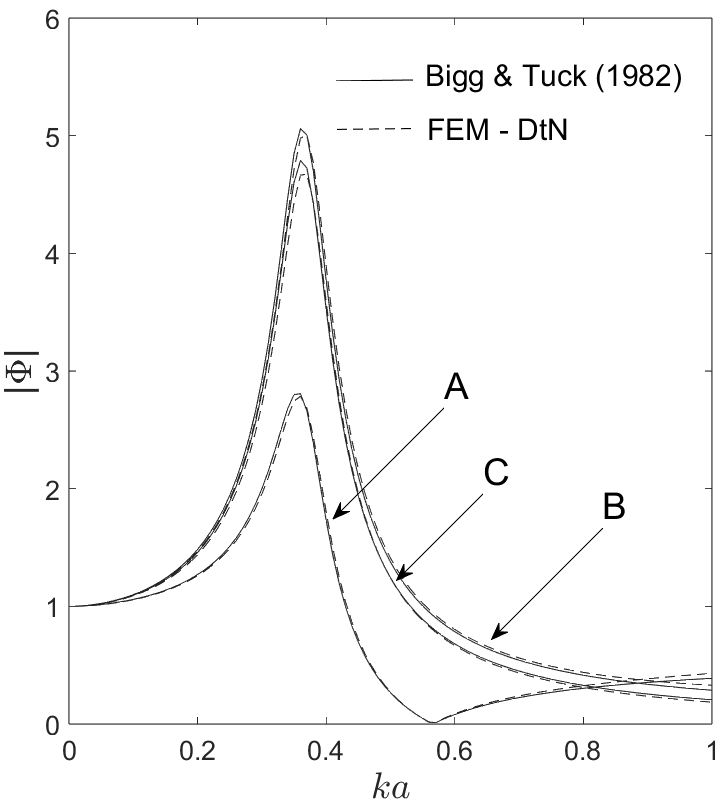}
    \end{tabular}
    }
    \caption{The circular harbour geometry with a small opening to the surrounding ocean, the mesh used (blue lines) and the sampled locations A, B and C (left). 
    The response spectra in terms of the modulus of the velocity potential versus nondimensional wavenumber $ka$ at locations A, B and C (right), using the code developed in this study (FEM with the DtN condition; solid curves) and by \citet{bigg1982two} (broken curves).}
    \label{fig:harbour}
\end{figure}

The code used in the study is modified for use on the degenerate problem of harbour excitation by long waves, in which the ice shelf (Kirchoff-Love plate) is removed from $\Omega$, which allows for verification against results from the existing literature.
The case of a circular harbour of radius $a$ was chosen, with a narrow opening to the surrounding ocean. 
\citet{bigg1982two} solved the problem using the method of matched asymptotoic expansions and gave results for the modulus of the velocity potential at three locations in the harbour (A, B and C in Figure~\ref{fig:harbour}, left) in their Figure~9.
The analogous results generated from the code used for the present study (using the DtN condition with $R=3a$ and $N_{\text{el}}=14,464$ elements) agree well with the results from \citet{bigg1982two} (Figure~\ref{fig:harbour}, right).

\begin{thebibliography}{63}
\providecommand{\natexlab}[1]{#1}
\providecommand{\url}[1]{\texttt{#1}}
\expandafter\ifx\csname urlstyle\endcsname\relax
  \providecommand{\doi}[1]{doi: #1}\else
  \providecommand{\doi}{doi: \begingroup \urlstyle{rm}\Url}\fi

\bibitem[Arthur et~al.(2021)Arthur, Stokes, Jamieson, Miles, Carr, and Leeson]{arthur2021triggers}
J.F. Arthur, C.R. Stokes, S.S. Jamieson, B.W. Miles, J.R. Carr, and A.A. Leeson.
\newblock The triggers of the disaggregation of voyeykov ice shelf (2007), wilkes land, east antarctica, and its subsequent evolution.
\newblock \emph{Journal of Glaciology}, 67\penalty0 (265):\penalty0 933--951, 2021.

\bibitem[Banwell et~al.(2017)Banwell, Willis, Macdonald, Goodsell, Mayer, Powell, and Macayeal]{banwell2017calving}
A.F. Banwell, I.C. Willis, G.J. Macdonald, B.~Goodsell, D.P. Mayer, A.~Powell, and D.R. Macayeal.
\newblock Calving and rifting on the mcmurdo ice shelf, antarctica.
\newblock \emph{Annals of Glaciology}, 58\penalty0 (75pt1):\penalty0 78--87, 2017.

\bibitem[Bassis and Walker(2012)]{bassis2012upper}
J.N Bassis and C.C Walker.
\newblock Upper and lower limits on the stability of calving glaciers from the yield strength envelope of ice.
\newblock \emph{Proceedings of the Royal Society A: Mathematical, Physical and Engineering Sciences}, 468\penalty0 (2140):\penalty0 913--931, 2012.

\bibitem[Bassis et~al.(2024)Bassis, Crawford, Kachuck, Benn, Walker, Millstein, Duddu, {\AA}str{\"o}m, Fricker, and Luckman]{bassis2024stability}
J.N. Bassis, A.~Crawford, S.B. Kachuck, D.I. Benn, C.~Walker, J.~Millstein, R.~Duddu, J.~{\AA}str{\"o}m, H.~Fricker, and A.~Luckman.
\newblock Stability of ice shelves and ice cliffs in a changing climate.
\newblock \emph{Annual Review of Earth and Planetary Sciences}, 52, 2024.

\bibitem[Bennetts(2025)]{bennetts_waves_2025}
L.G. Bennetts.
\newblock Waves in ice.
\newblock In \emph{Reference {Module} in {Earth} {Systems} and {Environmental} {Sciences}}, pages doi.org/10.1016/B978--0--323--85242--5.00037--3. Elsevier, 2025.
\newblock ISBN 978-0-12-409548-9.
\newblock \doi{10.1016/B978-0-323-85242-5.00037-3}.
\newblock URL \url{https://linkinghub.elsevier.com/retrieve/pii/B9780323852425000373}.

\bibitem[Bennetts and Meylan(2021)]{bennetts2021complex}
L.G. Bennetts and M.H. Meylan.
\newblock Complex resonant ice shelf vibrations.
\newblock \emph{SIAM Journal on Applied Mathematics}, 81\penalty0 (4):\penalty0 1483--1502, 2021.

\bibitem[Bennetts et~al.(2007)Bennetts, Biggs, and Porter]{bennetts2007multi}
LG~Bennetts, NRT Biggs, and D~Porter.
\newblock A multi-mode approximation to wave scattering by ice sheets of varying thickness.
\newblock \emph{Journal of Fluid Mechanics}, 579:\penalty0 413--443, 2007.

\bibitem[Bennetts et~al.(2024{\natexlab{a}})Bennetts, Williams, and Porter]{bennetts2024thin}
L.G. Bennetts, T.D. Williams, and R.~Porter.
\newblock A thin-plate approximation for ocean wave interactions with an ice shelf.
\newblock \emph{Journal of Fluid Mechanics}, 984:\penalty0 A48, 2024{\natexlab{a}}.

\bibitem[Bennetts et~al.(2024{\natexlab{b}})]{bennetts2024closing}
L.G. Bennetts et~al.
\newblock Closing the loops on southern ocean dynamics: From the circumpolar current to ice shelves and from bottom mixing to surface waves.
\newblock \emph{Reviews of Geophysics}, 62\penalty0 (3):\penalty0 e2022RG000781, 2024{\natexlab{b}}.

\bibitem[Bennetts and Liang(2026)]{bennetts2026modelling}
Luke~G Bennetts and Jie Liang.
\newblock Modelling dynamic strains on ice shelves resulting from flexural and extensional motions forced by ocean wave packets.
\newblock \emph{The ANZIAM Journal}, 68:\penalty0 e4, 2026.

\bibitem[Bigg and Tuck(1982)]{bigg1982two}
G.~R. Bigg and E.~O. Tuck.
\newblock Two-dimensional resonators with small openings.
\newblock \emph{The ANZIAM Journal}, 24\penalty0 (1):\penalty0 2--27, 1982.

\bibitem[Bromirski et~al.(2010)Bromirski, Sergienko, and MacAyeal]{bromirski2010transoceanic}
P.D. Bromirski, O.V. Sergienko, and D.R. MacAyeal.
\newblock Transoceanic infragravity waves impacting antarctic ice shelves.
\newblock \emph{Geophysical Research Letters}, 37\penalty0 (2), 2010.

\bibitem[Bromirski et~al.(2015)Bromirski, Diez, Gerstoft, Stephen, Bolmer, Wiens, Aster, and Nyblade]{bromirski2015ross}
P.D. Bromirski, A.~Diez, P.~Gerstoft, R.A. Stephen, T.~Bolmer, D.A. Wiens, R.C. Aster, and A.~Nyblade.
\newblock Ross ice shelf vibrations.
\newblock \emph{Geophysical Research Letters}, 42\penalty0 (18):\penalty0 7589--7597, 2015.

\bibitem[Bromirski et~al.(2017)Bromirski, Chen, Stephen, Gerstoft, Arcas, Diez, Aster, Wiens, and Nyblade]{bromirski2017tsunami}
P.D. Bromirski, Z.~Chen, R.A. Stephen, P.~Gerstoft, D.~Arcas, A.~Diez, R.C. Aster, D.A. Wiens, and A.~Nyblade.
\newblock Tsunami and infragravity waves impacting antarctic ice shelves.
\newblock \emph{Journal of Geophysical Research: Oceans}, 122\penalty0 (7):\penalty0 5786--5801, 2017.

\bibitem[Brunt et~al.(2011)Brunt, Okal, and MacAyeal]{brunt2011antarctic}
K.M. Brunt, E.A. Okal, and D.R. MacAyeal.
\newblock Antarctic ice-shelf calving triggered by the honshu (japan) earthquake and tsunami, march 2011.
\newblock \emph{Journal of Glaciology}, 57\penalty0 (205):\penalty0 785--788, 2011.

\bibitem[Cathles~IV et~al.(2009)Cathles~IV, Okal, and MacAyeal]{cathles2009seismic}
L.~M. Cathles~IV, E.~A. Okal, and D.~R. MacAyeal.
\newblock Seismic observations of sea swell on the floating {Ross Ice Shelf, Antarctica}.
\newblock \emph{Journal of Geophysical Research: Earth Surface}, 114\penalty0 (F2), 2009.

\bibitem[Chen et~al.(2018)Chen, Bromirski, Gerstoft, Stephen, Wiens, Aster, and Nyblade]{chen2018ocean}
Z.~Chen, P.~D. Bromirski, P.~Gerstoft, R.~A. Stephen, D.~A. Wiens, R.~C. Aster, and A.~A. Nyblade.
\newblock Ocean-excited plate waves in the {Ross and Pine Island Glacier} ice shelves.
\newblock \emph{Journal of Glaciology}, 64\penalty0 (247):\penalty0 730--744, 2018.

\bibitem[Chen et~al.(2019)Chen, Bromirski, Gerstoft, Stephen, Lee, Yun, Olinger, Aster, Wiens, and Nyblade]{chen2019ross}
Z.~Chen, P.~D. Bromirski, P.~Gerstoft, R.~A. Stephen, W.~S. Lee, S.~Yun, S.~D. Olinger, R.~C. Aster, D.~A. Wiens, and A.~A. Nyblade.
\newblock {Ross Ice Shelf} icequakes associated with ocean gravity wave activity.
\newblock \emph{Geophysical Research Letters}, 46\penalty0 (15):\penalty0 8893--8902, 2019.

\bibitem[Depoorter et~al.(2013)Depoorter, Bamber, Griggs, Lenaerts, Ligtenberg, van~den Broeke, and Moholdt]{depoorter2013calving}
M.A. Depoorter, J.L. Bamber, J.A. Griggs, J.T. Lenaerts, S.R. Ligtenberg, M.R. van~den Broeke, and G.~Moholdt.
\newblock Calving fluxes and basal melt rates of antarctic ice shelves.
\newblock \emph{Nature}, 502\penalty0 (7469):\penalty0 89--92, 2013.

\bibitem[Gomez-Fell et~al.(2022)Gomez-Fell, Rack, Purdie, and Marsh]{gomezfell2022parker}
R.~Gomez-Fell, W.~Rack, H.~Purdie, and O.~Marsh.
\newblock Parker ice tongue collapse, antarctica, triggered by loss of stabilizing land-fast sea ice.
\newblock \emph{Geophysical Research Letters}, 49\penalty0 (1):\penalty0 e2021GL096156, 2022.

\bibitem[Greene et~al.(2022)Greene, Gardner, Schlegel, and Fraser]{greene2022calving}
C.A. Greene, A.S. Gardner, N.-J. Schlegel, and A.D. Fraser.
\newblock Antarctic calving loss rivals ice-shelf thinning.
\newblock \emph{Nature}, 609\penalty0 (7929):\penalty0 948--953, 2022.

\bibitem[Holdsworth and Glynn(1978)]{holdsworth1978iceberg}
G.~Holdsworth and J.~Glynn.
\newblock Iceberg calving from floating glaciers by a vibrating mechanism.
\newblock \emph{Nature}, 274\penalty0 (5670):\penalty0 464--466, 1978.

\bibitem[Holdsworth and Glynn(1981)]{holdsworth1981mechanism}
G.~Holdsworth and J.E. Glynn.
\newblock A mechanism for the formation of large icebergs.
\newblock \emph{Journal of Geophysical Research: Oceans}, 86\penalty0 (C4):\penalty0 3210--3222, 1981.

\bibitem[Ilyas et~al.(2018)Ilyas, Meylan, Lamichhane, and Bennetts]{ilyas2018timedomain}
M.~Ilyas, M.H. Meylan, B.~Lamichhane, and L.G. Bennetts.
\newblock Time-domain and modal response of ice shelves to wave forcing using the finite element method.
\newblock \emph{Journal of Fluids and Structures}, 80:\penalty0 113--131, 2018.

\bibitem[Isaacson and Qu(1990)]{isaacson1990waves}
Michael Isaacson and Shiqin Qu.
\newblock Waves in a harbour with partially reflecting boundaries.
\newblock \emph{Coastal Engineering}, 14\penalty0 (3):\penalty0 193--214, 1990.

\bibitem[Kalyanaraman et~al.(2019)Kalyanaraman, Bennetts, Lamichhane, and Meylan]{kalyanaraman2019shallow}
B.~Kalyanaraman, L.G. Bennetts, B.~Lamichhane, and M.H. Meylan.
\newblock On the shallow-water limit for modelling ocean-wave induced ice-shelf vibrations.
\newblock \emph{Wave Motion}, 90:\penalty0 1--16, 2019.

\bibitem[Kalyanaraman et~al.(2020)Kalyanaraman, Meylan, Bennetts, and Lamichhane]{kalyanaraman2020coupled}
B.~Kalyanaraman, M.H. Meylan, L.G. Bennetts, and B.P. Lamichhane.
\newblock A coupled fluid-elasticity model for the wave forcing of an ice-shelf.
\newblock \emph{Journal of Fluids and Structures}, 97:\penalty0 103074, 2020.

\bibitem[Kalyanaraman et~al.(2021)Kalyanaraman, Meylan, Lamichhane, and Bennetts]{kalyanaraman2021icefem}
B.~Kalyanaraman, M.H. Meylan, B.P. Lamichhane, and L.G. Bennetts.
\newblock icefem: A freefem package for wave induced ice-shelf vibrations.
\newblock \emph{Journal of Open Source Software}, 6\penalty0 (59), 2021.

\bibitem[Karperaki et~al.(2019)Karperaki, Papathanasiou, and Belibassakis]{karperaki2019optimized}
A.E. Karperaki, T.K. Papathanasiou, and K.A. Belibassakis.
\newblock An optimized, parameter-free pml-fem for wave scattering problems in the ocean and coastal environment.
\newblock \emph{Ocean Engineering}, 179:\penalty0 307--324, 2019.

\bibitem[Li et~al.(2021)Li, Shi, and Wu]{li2021interaction}
ZF~Li, YY~Shi, and GX~Wu.
\newblock Interaction of ocean wave with a harbor covered by an ice sheet.
\newblock \emph{Physics of Fluids}, 33\penalty0 (5), 2021.

\bibitem[Liang et~al.(2024)Liang, Pitt, and Bennetts]{liang2024pan}
J.~Liang, J.P. Pitt, and L.G. Bennetts.
\newblock Pan‐antarctic assessment of ice shelf flexural responses to ocean waves.
\newblock \emph{Journal of Geophysical Research: Oceans}, 129\penalty0 (8):\penalty0 e2023JC020824, 2024.

\bibitem[Lipovsky(2018)]{lipovsky2018ice}
B.~P. Lipovsky.
\newblock Ice shelf rift propagation and the mechanics of wave-induced fracture.
\newblock \emph{Journal of Geophysical Research: Oceans}, 123\penalty0 (6):\penalty0 4014--4033, 2018.

\bibitem[MacAyeal et~al.(2015)MacAyeal, Sergienko, and Banwell]{macayeal2015model}
Douglas~R MacAyeal, Olga~V Sergienko, and Alison~F Banwell.
\newblock A model of viscoelastic ice-shelf flexure.
\newblock \emph{Journal of Glaciology}, 61\penalty0 (228):\penalty0 635--645, 2015.

\bibitem[MacAyeal et~al.(2006)MacAyeal, Okal, Aster, Bassis, Brunt, Cathles, Drucker, Fricker, Kim, Martin, and Okal]{macayeal2006transoceanic}
D.R. MacAyeal, E.A. Okal, R.C. Aster, J.N. Bassis, K.M. Brunt, L.M. Cathles, R.~Drucker, H.A. Fricker, Y.J. Kim, S.~Martin, and M.H. Okal.
\newblock Transoceanic wave propagation links iceberg calving margins of antarctica with storms in tropics and northern hemisphere.
\newblock \emph{Geophysical Research Letters}, 33\penalty0 (17), 2006.

\bibitem[Massom et~al.(2018)Massom, Scambos, Bennetts, Reid, Squire, and Stammerjohn]{massom2018disintegration}
R.A. Massom, T.A. Scambos, L.G. Bennetts, P.~Reid, V.A. Squire, and S.E. Stammerjohn.
\newblock Antarctic ice shelf disintegration triggered by sea ice loss and ocean swell.
\newblock \emph{Nature}, 558\penalty0 (7710):\penalty0 383--389, 2018.

\bibitem[Mei et~al.(2005)Mei, Stiassnie, and Yue]{mei2005theory}
C~C Mei, M~A Stiassnie, and D~K-P Yue.
\newblock \emph{Theory and applications of ocean surface waves: Part 1: linear aspects}.
\newblock World Scientific, 2005.

\bibitem[Melenk and Sauter(2010)]{melenk2010convergence}
J.~Melenk and S.~Sauter.
\newblock Convergence analysis for finite element discretizations of the helmholtz equation with dirichlet-to-neumann boundary conditions.
\newblock \emph{Mathematics of Computation}, 79\penalty0 (272):\penalty0 1871--1914, 2010.

\bibitem[Meylan et~al.(2017)Meylan, Bennetts, Hosking, and Catt]{meylan2017normal}
M.H. Meylan, L.G. Bennetts, R.J. Hosking, and E.~Catt.
\newblock On the calculation of normal modes of a coupled ice-shelf/sub-ice-shelf cavity system.
\newblock \emph{Journal of Glaciology}, 63\penalty0 (240):\penalty0 751--754, 2017.

\bibitem[Meylan et~al.(2021)Meylan, Ilyas, Lamichhane, and Bennetts]{meylan2021swell}
M.H. Meylan, M.~Ilyas, B.P. Lamichhane, and L.G. Bennetts.
\newblock Swell-induced flexural vibrations of a thickening ice shelf over a shoaling seabed.
\newblock \emph{Proceedings of the Royal Society A}, 477\penalty0 (2254):\penalty0 20210173, 2021.

\bibitem[Mitsoudis et~al.(2012)Mitsoudis, Makridakis, and Plexousakis]{mitsoudis2012helmholtz}
Dimitrios~A Mitsoudis, Ch~Makridakis, and Michael Plexousakis.
\newblock Helmholtz equation with artificial boundary conditions in a two-dimensional waveguide.
\newblock \emph{SIAM Journal on Mathematical Analysis}, 44\penalty0 (6):\penalty0 4320--4344, 2012.

\bibitem[Noble et~al.(2020)]{noble2020sensitivity}
T.L. Noble et~al.
\newblock The sensitivity of the antarctic ice sheet to a changing climate: Past, present, and future.
\newblock \emph{Reviews of Geophysics}, 58:\penalty0 e2019RG000663, 2020.

\bibitem[Oberai et~al.(1998)Oberai, Malhotra, and Pinsky]{oberai1998implementation}
A.A. Oberai, M.~Malhotra, and P.M. Pinsky.
\newblock On the implementation of the dirichlet-to-neumann radiation condition for iterative solution of the helmholtz equation.
\newblock \emph{Applied Numerical Mathematics}, 27:\penalty0 443--464, 1998.

\bibitem[Ochwat et~al.(2023)Ochwat, Scambos, Banwell, Anderson, Maclennan, Picard, Shates, Marinsek, Margonari, Truffer, and Pettit]{ochwat2023triggers}
N.E. Ochwat, T.A. Scambos, A.F. Banwell, R.S. Anderson, M.L. Maclennan, G.~Picard, J.A. Shates, S.~Marinsek, L.~Margonari, M.~Truffer, and E.C. Pettit.
\newblock Triggers of the 2022 larsen b multi-year landfast sea ice break-out and initial glacier response.
\newblock \emph{The Cryosphere Discussions}, 2023:\penalty0 1--34, 2023.

\bibitem[Papathanasiou and Belibassakis(2018)]{papathanasiou2018resonances}
T.K. Papathanasiou and K.A. Belibassakis.
\newblock Resonances of enclosed shallow water basins with slender floating elastic bodies.
\newblock \emph{Journal of Fluids and Structures}, 82:\penalty0 538--558, 2018.

\bibitem[Papathanasiou et~al.(2019)Papathanasiou, Karperaki, and Belibassakis]{papathanasiou2019hydroelastic}
T.K. Papathanasiou, A.E. Karperaki, and K.A. Belibassakis.
\newblock On the resonant hydroelastic behaviour of ice shelves.
\newblock \emph{Ocean Modelling}, 133:\penalty0 11--26, 2019.

\bibitem[Papathanasiou et~al.(2025)Papathanasiou, Bennetts, and Meylan]{papathanasiou2025modelling}
T.K. Papathanasiou, L.G. Bennetts, and M.H. Meylan.
\newblock Modelling hydroelastic flexure of arbitrarily shaped ice shelves forced by long ocean waves.
\newblock \url{https://doi.org/10.17632/jvmt67xpb2.1}, 2025.
\newblock Mendeley Data, V1.

\bibitem[Rannacher(1979)]{rannacher1979nonconforming}
R.~Rannacher.
\newblock Nonconforming finite element methods for eigenvalue problems in linear plate theory.
\newblock \emph{Numerische Mathematik}, 33:\penalty0 23--42, 1979.

\bibitem[Robinson and Haskell(1992)]{robinson1992travelling}
WH~Robinson and TG~Haskell.
\newblock Travelling flexural waves in the {Erebus Glacier Tongue, McMurdo Sound, Antarctica}.
\newblock \emph{Cold Regions Science and Technology}, 20\penalty0 (3):\penalty0 289--293, 1992.

\bibitem[Scambos et~al.(2013)Scambos, Hulbe, and Fahnestock]{scambos2013climate}
T.~Scambos, C.~Hulbe, and M.~Fahnestock.
\newblock Climate-induced ice shelf disintegration in the antarctic peninsula.
\newblock In \emph{Antarctic Peninsula Climate Variability: Historical and Paleoenvironmental Perspectives}, pages 79--92. 2013.

\bibitem[Sergienko(2010)]{sergienko2010elastic}
O.V. Sergienko.
\newblock Elastic response of floating glacier ice to impact of long‐period ocean waves.
\newblock \emph{Journal of Geophysical Research: Earth Surface}, 115\penalty0 (F4), 2010.

\bibitem[Sergienko(2013)]{sergienko2013normal}
O.V. Sergienko.
\newblock Normal modes of a coupled ice-shelf/sub-ice-shelf cavity system.
\newblock \emph{Journal of Glaciology}, 59\penalty0 (213):\penalty0 76--80, 2013.

\bibitem[Sergienko(2017)]{sergienko2017behavior}
O.V. Sergienko.
\newblock Behavior of flexural gravity waves on ice shelves: Application to the {Ross Ice Shelf}.
\newblock \emph{Journal of Geophysical Research: Oceans}, 122\penalty0 (8):\penalty0 6147--6164, 2017.

\bibitem[Specht(1988)]{specht1988modified}
B.~Specht.
\newblock Modified shape functions for the three‐node plate bending element passing the patch test.
\newblock \emph{International Journal for Numerical Methods in Engineering}, 26\penalty0 (3):\penalty0 705--715, 1988.

\bibitem[Squire et~al.(1994)Squire, Robinson, Meylan, and Haskell]{squire1994observations}
V~A Squire, W~H Robinson, M~H Meylan, and T~G Haskell.
\newblock Observations of flexural waves on the {Erebus Ice Tongue, McMurdo Sound, Antarctica}, and nearby sea ice.
\newblock \emph{Journal of Glaciology}, 40\penalty0 (135):\penalty0 377--385, 1994.

\bibitem[Steward and Panchang(2001)]{steward2001improved}
David~R Steward and Vijay~G Panchang.
\newblock Improved coastal boundary condition for surface water waves.
\newblock \emph{Ocean Engineering}, 28\penalty0 (1):\penalty0 139--157, 2001.

\bibitem[Tazhimbetov et~al.(2023)Tazhimbetov, Almquist, Werpers, and Dunham]{tazhimbetov2023simulation}
N.~Tazhimbetov, M.~Almquist, J.~Werpers, and E.M. Dunham.
\newblock Simulation of flexural-gravity wave propagation for elastic plates in shallow water using an energy-stable finite difference method with weakly enforced boundary and interface conditions.
\newblock \emph{Journal of Computational Physics}, 493:\penalty0 112470, 2023.

\bibitem[Teder et~al.(2025)Teder, Bennetts, Reid, Massom, Pitt, Scambos, and Fraser]{teder2025large}
Nathan~J Teder, Luke~G Bennetts, Phillip~A Reid, Robert~A Massom, Jordan~PA Pitt, Theodore~A Scambos, and Alexander~D Fraser.
\newblock Large-scale ice-shelf calving events follow prolonged amplifications in flexure.
\newblock \emph{Nature Geoscience}, 18\penalty0 (7):\penalty0 599--606, 2025.

\bibitem[Teder et~al.(2022)Teder, Bennetts, Reid, and Massom]{teder2022sea}
N.J. Teder, L.G. Bennetts, P.A. Reid, and R.A. Massom.
\newblock Sea ice-free corridors for large swell to reach {A}ntarctic ice shelves.
\newblock \emph{Environmental Research Letters}, 17\penalty0 (4):\penalty0 045026, 2022.

\bibitem[Thiel et~al.(1960)Thiel, Crary, Haubrich, and Behrendt]{thiel1960gravimetric}
E~Thiel, A~P Crary, R~A Haubrich, and J~C Behrendt.
\newblock Gravimetric determination of ocean tide, {Weddell and Ross Seas, Antarctica}.
\newblock \emph{Journal of Geophysical Research}, 65\penalty0 (2):\penalty0 629--636, 1960.

\bibitem[Timoshenko and Woinowsky-Krieger(1959)]{timoshenko1959theory}
S.S. Timoshenko and S.~Woinowsky-Krieger.
\newblock \emph{Theory of Plates and Shells}.
\newblock McGraw Hill, New York, 1959.

\bibitem[Vinogradov and Holdsworth(1985)]{vinogradov1985oscillation}
O.G. Vinogradov and G.~Holdsworth.
\newblock Oscillation of a floating glacier tongue.
\newblock \emph{Cold Regions Science and Technology}, 10\penalty0 (3):\penalty0 263--271, 1985.

\bibitem[Williams and Robinson(1981)]{williams1981flexural}
RT~Williams and ES~Robinson.
\newblock Flexural waves in the {Ross Ice Shelf}.
\newblock \emph{Journal of Geophysical Research: Oceans}, 86\penalty0 (C7):\penalty0 6643--6648, 1981.

\bibitem[Zhao et~al.(2024)Zhao, Cheng, Fraser, Bennetts, Xiao, Liang, Li, and Lia]{zhao2024longterm}
A.~Zhao, Y.~Cheng, A.D. Fraser, L.G. Bennetts, X.~Xiao, Q.~Liang, T.~Li, and R.~Lia.
\newblock Long-term evolution of the sulzberger ice shelf, west antarctica: Insights from 74-year observations and 2022 hunga-tonga volcanic tsunami-induced calving.
\newblock \emph{Earth and Planetary Science Letters}, 646:\penalty0 118958, 2024.

\end{thebibliography}





\end{document}